 \documentclass[final,3p]{elsarticle}

 \usepackage{graphics}
\usepackage{graphicx}

\usepackage{amssymb}
\usepackage{amsthm}
\usepackage{amsmath}
\usepackage{hyperref}

\newcounter{bla}

\def\SUSEFLAV{{\tt SuSeFLAV}}
\def\SOFTSUSY{{\tt SOFTSUSY}}
\def\SUSPECT{{\tt SuSpect}}
\def\LAPACK{{\tt LAPACK}}

\def\ISASUSY{{\tt ISASUSY}}

\def\MICROMEGAS{{\tt MicrOMEGAs}}
\def\DARKSUSY{{\tt DarkSUSY}}
\newcommand{\spheno}{\texttt{SPheno}}
\def\SUPERISO{{\tt SuperIso}}

\def\half{\frac{1}{2}}
\def\onethrd{\frac{1}{3}}
\def\twothrd{\frac{2}{3}}
\def\onesix{\frac{1}{6}}

\def\beq{\begin{equation}}
\def\eeq{\end{equation}}

\def\ifmath#1{\relax\ifmmode #1\else $#1$\fi}

\def\Y{\ifmath{{\bf Y}}}

\def\A{\ifmath{{\bf A}}}

\def\Y{{\bf Y}}
\def\m{{\bf m}}
\def\A{{\bf A}}
\def\M{{\bf m}}

\newcommand{\tb}{\tan\beta}

\newcommand{\mgut}{M_{\rm GUT}}
\newcommand{\msusy}{M_{\rm SUSY}}
\newcommand{\W}{W^{+}}
\newcommand{\Hp}{H^{+}}

\newcommand{\qL}{\tilde{q}_{L}}
\newcommand{\qR}{\tilde{q}_{R}}
\newcommand{\seL}{\tilde{l}_{L}}
\newcommand{\seR}{\tilde{l}_{R}}
\newcommand{\nuL}{\tilde{\nu}_{l}}

\newcommand{\quL}{\tilde{u}_{L}}
\newcommand{\quR}{\tilde{u}_{R}}
\newcommand{\qdL}{\tilde{d}_{L}}
\newcommand{\qdR}{\tilde{d}_{R}}

\newcommand{\stL}{\tilde{t}_{1}}
\newcommand{\stR}{\tilde{t}_{2}}
\newcommand{\sbL}{\tilde{b}_{1}}
\newcommand{\sbR}{\tilde{b}_{2}}
\newcommand{\stauL}{\tilde{\tau}_{1}}
\newcommand{\stauR}{\tilde{\tau}_{2}}
\newcommand{\nutau}{\tilde{\nu}_{\tau}}

\newcommand{\neut}[1]{\tilde{\chi}^{0}_{#1}}
\newcommand{\charg}[1]{\tilde{\chi}^{\pm}_{#1}}

\journal{Computer Physics Communications}

\begin{document}

\begin{frontmatter}



\title{\SUSEFLAV\ 1.2: program for supersymmetric mass spectra with seesaw mechanism and rare lepton flavor violating decays}

\author[deb]{Debtosh Chowdhury\corref{author1}}
\address[deb]{Centre for High Energy Physics,Indian Institute of Science, Bangalore 560 012, India}
\author[deb,rag]{Raghuveer Garani\corref{author2}}
\address[rag]{Physikalisches Institut der Universit{\"a}t Bonn,  Nu\ss alle 12, D-53115  Bonn, Germany}
\author[deb]{Sudhir K. Vempati\corref{author3}}
\cortext[author1] {\textit{E-mail address:} debtosh@cts.iisc.ernet.in}
\cortext[author2] {\textit{E-mail address:} s6ragara@uni-bonne.de}
\cortext[author3] {\textit{E-mail address:} vempati@cts.iisc.ernet.in}

\begin{abstract}
Accurate supersymmetric spectra are required to confront data from direct and indirect searches of supersymmetry. \SUSEFLAV\footnote{\url{http://cts.iisc.ernet.in/Suseflav/main.html}}\ is a numerical tool which is capable of computing supersymmetric spectra precisely for various supersymmetric breaking scenarios  applicable even in the presence of flavor violation. The program solves MSSM RGEs with complete $3\times3$ flavor mixing at 2-loop level and one loop finite threshold corrections to all  MSSM parameters by incorporating radiative electroweak symmetry breaking conditions. The program also incorporates the Type-I seesaw mechanism  with  three massive right handed neutrinos at user defined  mass scales and  mixing. It also computes branching ratios of flavor violating processes such  as $l_j\,\rightarrow\, l_i\gamma$,  $l_j\;\rightarrow\, 3 ~l_i$,  $b \,\rightarrow\,s\gamma$ and supersymmetric contributions to flavor conserving quantities such as $(g_{\mu}-2)$. A large choice of executables suitable for various operations of the program are provided.
\end{abstract}

\begin{keyword}
MSSM \sep Right Handed Neutrinos \sep Lepton Flavor Violation.

\end{keyword}

\end{frontmatter}



\noindent
{\bf PROGRAM SUMMARY} \\

\noindent
{\em Manuscript Title: \SUSEFLAV:  program for supersymmetric mass spectra with seesaw mechanism and rare lepton flavor violating decays}                                       \\
{\em Authors: Debtosh Chowdhury, Raghuveer Garani, Sudhir K. Vempati}                                                \\
{\em Program Title: \SUSEFLAV}                                          \\
{\em Journal Reference:}                                      \\
{\em Catalogue identifier:}                                   \\
{\em Licensing provisions: GNU Public License}                               \\
{\em Programming language: Fortran 95}                                   \\
{\em Computer: Personal Computer, Work-Station}                                               \\
{\em Operating system: Linux, Unix}                                       \\
{\em RAM: }                                               \\
{\em Number of processors used: Single}                              \\
{\em Supplementary material:}                                 \\
{\em Keywords: MSSM, Right Handed Neutrinos, Lepton Flavor Violation.}  \\
{\em Classification:}                                         \\
{\em External routines/libraries:} \\ 
{\em Subprograms used: } \\ 
{\em Nature of problem: Determination of masses and mixing of supersymmetric particles within the context of MSSM with conserved R-parity with and without the  presence of Type-I seesaw. Inter-generational mixing is considered while calculating the mass spectrum. Supersymmetry breaking parameters are taken as  inputs at a high scales specified by  the mechanism of supersymmetry breaking. RG equations including full inter-generational mixing are then used to evolve these parameters up to the electroweak breaking scale. The low energy  supersymmetric spectrum is calculated at the scale where successful radiative electro-weak symmetry breaking occurs. At weak scale Standard Model fermion masses, gauge couplings are determined ncluding the supersymmetric radiative corrections. Once the spectrum is computed,  the program proceeds to various lepton flavor violating observables ({\it e.g.}, BR$(\mu\ \rightarrow\ e \gamma)$, BR$(\tau\ \rightarrow\ \mu \gamma)$ etc.) at the weak scale. } \\
   \\
{\em Solution method: Two loop RGEs with full $3 \times 3$ flavor mixing for all supersymmetry breaking parameters are used to compute the low energy supersymmetric mass spectrum. 
Adaptive step size Runge-Kutta method is used to solve the RGEs numerically between the high scale and the electroweak breaking scale. Iterative procedure is employed to get the consistent radiative electroweak symmetry breaking condition. 
The masses of the supersymmetric particles are computed at 1-loop order. The third generation SM particles and the gauge couplings are evaluated at the 1-loop  order including supersymmetric corrections.  A further iteration of the full program is employed such that the SM masses and couplings are consistent with the supersymmetric particle spectrum. }\\
   \\
{\em Restrictions:}\\
   \\
{\em Unusual features:}\\ 
   \\
{\em Additional comments:}\\
Several executables are presented for the user. 
   \\
{\em Running time:  0.2 seconds on a Intel(R) Core(TM) i5 CPU 650 with 3.20 GHz.}\\
   \\


\section{Introduction}

Low energy supersymmetry \cite{martin} is currently being probed by the Large Hadron Collider (LHC) at CERN and the Tevatron collider at the Fermilab\footnote{The Tevatron collider has recently stopped functioning.}.  On the other hand, there is already a huge amount of information which has been collected and is being collected which gives information on low energy supersymmetric Lagrangian indirectly. For example,  the flavor experiments  in the hadronic and the leptonic sectors  place strong constraints on the flavor off-diagonal entries in the Lagrangian. Similarly, the astrophysical data on dark matter which  has been improved with the latest WMAP 7-year results \cite{Larson:2010gs} also strongly restricts the parameter space where the mass and the couplings of the lightest supersymmetric particle (LSP) correspond to the observed relic density. While experimental evidence for supersymmetry is definitely far more superior compared to the indirect detection of supersymmetry, the power of indirect experimental data to constrain the parameter space cannot be underestimated.  Furthermore, as it is well known even if there is a positive experimental signal at the LHC, it would be very hard to reconstruct the supersymmetric breaking Lagrangian unambiguously due to the large number of degeneracies present in the parameter space which can give similar signals at the colliders \cite{arkani-kane}. It has also been noted that flavor violating observables and dark matter could help to break these degeneracies \cite{berger}.  

Obviously, the flavor observables depend on the supersymmetric model in which they are calculated. Unfortunately, most of the present supersymmetric mass spectrum calculators do not take in to consideration the effect of flavor violation in the running of soft mass parameters either in the hadronic sector or the leptonic 
sector\footnote{The recent version of {\tt SPheno} \cite{spheno2} is an exception.}. In hadronic sector, typically the CKM is considered to be the only source of flavor violation,  while this works very well, as long as one restricts to the scheme of Minimal Flavor Violation (MFV), in a more general scheme of supersymmetric models, such an assumption cannot be supported. Recently `observed' deviations from the Standard Model CKM paradigm \cite{soni,buras}
might find explanations in terms of a  supersymmetric standard model with some amount of flavor violation \cite{Silvestrini:2007yf}. To study the associated phenomenology of such kinds of models either for dark matter relic density, collider searches or
other theoretical aspects such as threshold corrections to fermion masses, gauge coupling unification etc., would require precise computation of the mass spectrum in these models.

In the leptonic sector, the case for flavor violation is even more stronger. Firstly, neutrinos have non-zero masses and secondly their flavor mixing is large as has been observed in the neutrino oscillation experiments.  Most of mechanisms
of generating neutrino masses and mixing inevitably lead to large flavor violation in supersymmetric theories. One of the simplest ways to generate neutrino masses is the so called `seesaw mechanism'. In the present work, we have restricted ourselves to Type-I seesaw mechanism, though the program can be generalized to incorporate other seesaw mechanisms by adding the corresponding RGEs.  The Type-I seesaw mechanism\footnote{For a summary of the seesaw mechanisms, please see \cite{davidson}.} has three singlet right handed neutrinos added to the Standard Model which leads to two additional terms, the Dirac mass term combining the left and right neutrino fields and the lepton number violating Majorana mass term for the right handed singlet fields.  The interplay between these two terms leads to small  Majorana masses to the left handed neutrinos in the limit of heavy Majorana masses for the right neutrinos.  The supersymmetric version of the seesaw mechanism  was proposed long ago \cite{bm86} and some of its consequences for leptonic flavor violations have  immediately been noticed.  Over the years, other 
theoretical/phenomenological consequences of having right handed neutrinos has been observed.  In the following we list some of them. 

\begin{itemize}
 \item
 {\bf $Y_{b}-Y_{\tau}$ Unification:}  The presence of right handed neutrinos could significantly modify regions where $\tau-b$ Yukawa couplings
 unify at the GUT scale in unified theories like SO(10) or SU(5) \cite{andrea, smirnov,yanagida,gomez}. This is due to the fact that the neutrino 
 Dirac Yukawa couplings enter the renormalization group equations (RGE) of the $Y_{b}$ and $Y_{\tau}$ at 2-loop level and $Y_{t}$ at the 1-loop.  This is enough to change the
 $Y_{b}/Y_{\tau}$ ratio at the weak scale, if the neutrino Yukawa couplings are large. 
 
 \item 
 {\bf Lepton Flavor Violation:} As mentioned previously, one of the main consequences of the seesaw mechanism in supersymmetric theories is the 
 violation of lepton numbers leading to rare flavor violating decays \cite{bm86}. This flavor violation will be generated through the RGE 
 even if the supersymmetry  breaking mechanism at the high scale  conserves flavor as in mSUGRA. In particular GUT models, the 
 generated flavor violations could be large enough to strongly constrain observability of supersymmetry at the LHC \cite{Hisano,calibbi,ibarra,paride}.
 On going experiments like MEG and future experiments like PRISM/PRIME and Super-B factories have enhanced sensitivity to large amounts of
 parameter space even with small mixing and small $\tan\beta$ \cite{calibbi,paride}.
 
 \item 
 {\bf Dark Matter:} One of the most surprising phenomenological aspects of seesaw mechanism and SUSY-GUT models has been the impact
 on Dark matter phenomenology.  The presence of a single right handed neutrino with a large Yukawa coupling could significantly enhance
 the efficiency of the electroweak symmetry breaking and thus making the focus point region unviable within mSUGRA like models \cite{cmv,barger}.
 Similarly, GUT effects can significantly effect the stau co-annihilation region \cite{cmv}.  The co-annihilation region and the focus point regions
 seem to be most vulnerable to these effects  in other GUT models and  mSUGRA incorporating Type-II or Type-III seesaw mechanisms \cite{others}. 
 It has been recently realized that even flavor effects could play a role in the relic density calculations in the early universe \cite{fc}.
 
  \item
 {\bf Hadronic flavor violation:} 
 Incorporating Type-I seesaw mechanism in Grand Unified Theories (GUT) also has effects on the hadronic sector. For example, CP violation in the neutrino sector could be transmitted to the quark sector in SU(5) or SO(10) theories \cite{moroi,cmm}. The large phases of the neutrino mass matrix can be transmitted to the hadronic sector with effects in $K$ and $B$ physics phenomenology.  More generally in GUT theories, hadronic and leptonic flavor violations are related to each other by the underlying
 GUT symmetry \cite{luca}. 
 
  \item
{\bf Collider Signals:} 
Lepton flavor violation which might be inevitable due to the presence of a seesaw mechanism, can lead to flavor violation in the sleptonic sector as we have mentioned above.  Such flavor violation can be studied at the colliders by measuring the mass differences between the sleptons by observing such as sleptonic oscillations etc. \cite{arkani-hamed, hisanonojiri, calibbiburas, porodgroup}.

 \item
 {\bf Gauge Coupling Unification:} 
Finally, let us note that it has been pointed that the presence of massive right handed neutrinos with large yukawa couplings in the MSSM Lagrangian improves 
the accurate unification of the gauge couplings because $\alpha_{3}$ gets contribution from the right handed neutrinos through the RGE running
 at the two loop level \cite{ibarra-gaugecoupling}. 

\end{itemize}
 The above points provide enough justification to determine the supersymmetric mass spectrum  in seesaw models at a high precision level including the effects
  due to flavor violation\footnote{Programs like SuperIso\cite{superiso} and SUSYFLAVOR\cite{susyflavour}  compute the hadronic  flavor violating processes
   for the given supersymmetric spectrum at low energy
  with high precision.} . Unfortunately, while there exist very good spectrum calculators for supersymmetric theories like \ISASUSY\ \cite{isasusy}, \SUSPECT\ \cite{suspect}
and \SOFTSUSY\ \cite{softsusy}, they do not consider full flavor violating structure  in the computation of the soft spectrum either in their RGE's or their mass matrices\footnote{ However there are programs which consider flavor violation in their mass matrices. The program SPICE \cite{spice} computes mass matrices with full flavor violation without considering intergenerational mixing
in the RGE. After this work has appeared we have been informed that \SOFTSUSY\ also has a new version where flavor violation is considered \cite{softsusy_fv}.}. For these reasons, present versions of these programs might not be suitable for attacking problems listed above unless one significantly modifies them. 
Our program was written to address this deficiency in publicly available codes.  We, however point out  that the recent version 
of {\tt SPheno} \cite{spheno} is very similar to the program we are presenting here. It has full flavor structure for the soft masses as well as Yukawas at the 2-loop level and 
considers the full $6 \times 6$ mass matrices for the sparticles.    
The preliminary version of our program was first presented at \cite{daesymposium}. The present version is an expanded and more structured version of the same. This paper explains the code in detail. We attempted to link the physics discussion with the file structure of the code wherever possible such that the user can modify the code with minimal efforts. 

\SUSEFLAV\ is a program written in {\tt Fortran 95} in a fixed length format. The recommended compiler for this program is {\tt gfortran} available in various distributions of Linux. The program can be made executable using other Fortran compilers too, such as {\tt ifort},  by modifying the {\tt Makefile} in the main directory.  In addition to studying the spectrum of supersymmetric particles the code also computes leptonic flavor violating decays and some hadronic decays. We have implemented the {\tt SLHA 2.2} \cite{slha} format for dealing with the input parameters and output data. This way, the output of the code can be fed in to other publicly available programs either for computing Dark Matter relic density or for calculating supersymmetric particle decays and production cross-sections at LHC etc. While the main set of RGE's are written for Type-I seesaw mechanism, extending the program for other seesaw mechanisms or even other models would require coding the RGE's from respective models. However, other parts of the code, like mass spectrum, one-loop corrections etc., which are given in separate files in the source directory can be used to suit the user's needs. 

The rest of the paper is organized as follows. In section \ref{mssmrn} we describe the MSSM Lagrangian with Type-I seesaw mechanism. In section \ref{susybreak}, we discuss the various supersymmetry breaking scenarios considered in \SUSEFLAV.  In section \ref{algorithm} we describe the calculation of the low energy  supersymmetric spectrum implemented in the program. In section \ref{constraints} we briefly describe about various theoretical constraints we impose on the sparticle spectrum in \SUSEFLAV.  In section \ref{observables} we discuss about the various low energy observables computed in the program. In section \ref{exec} we show the instructions on how to install and execute \SUSEFLAV. In \ref{app:tree} we write down all the tree level mass matrices of the sparticles and in \ref{oneloop} we outline all the one-loop radiative corrections to these parameters \SUSEFLAV. In \ref{prog_flow} we graphically show the file structure in \SUSEFLAV. We close with a comparison chart of sparticle masses with other available codes in \ref{comp}. 


\section{Minimal Supersymmetric Seesaw Model}
\label{mssmrn}
Since 1998, ever increasing data from neutrino sector has firmly established that neutrinos have non-zero masses and that their flavors mix with two of their angles being close to maximal and the third angle is non-zero \cite{neutrino_review}. One of the elegant mechanisms to generate non-zero neutrino masses is  through the seesaw mechanism \cite{seesaw}, where right handed singlet fields are added to the Standard Model particle spectrum. These singlet neutrinos break lepton number typically at a scale much larger than the standard model scale through their  Majorana masses. 

Supersymmetric version of the seesaw mechanism is straight forward extension of the Minimal Supersymmetric Standard Model \cite{martin} 
by adding right handed neutrino superfields. The field content of the MSSMRN(MSSM $+$ Right Handed Neutrinos) and their transformation
properties under the gauge group ${\mathcal G}_{{\rm SM}}\equiv {\rm SU} (3)_c\otimes {\rm SU}(2)_{L}\otimes {\rm U}(1)_{Y}$  is given as 
\begin{eqnarray}
{L}&:&\left(1,2,-\frac{1}{2} \right);\quad {e^c}:\left(1,1,+1\right);\qquad {\nu^c}:\left(1,1,0\right); \nonumber \\
Q&:&\left(3,2,+\onesix \right);\quad {u^c}:\left({\bar 3},1,-\twothrd \right);\quad {d^c}:\left({\bar 3},1,\onethrd \right);\\
H_u&:&\left(1,2,+\half \right); \quad  H_d:\left(1,2,-\half \right). \nonumber 
\label{fields}
\end{eqnarray}
where $Q$ and $L$ stand for the $SU(2)_{L}$ doublet quarks and leptons,  $u^c$, $d^c$, $e^c$ and $\nu^c$ stand  for the $SU(2)_{L}$ singlet 
quarks and leptons, and the two Higgs doublet chiral superfields are denoted by  $H_u$ and $H_d$.  At the seesaw scale and above, $q^2 \gtrsim M_{R}^2$, 
the superpotential  takes the form : 
\begin{align}
\mathcal{W}\  =\ &
\Y^d_{ij} d^c_i Q_j  H_d +  \Y^u_{ij} u^c_i Q_j  H_u +
\Y^e_{ij} e^c_i  L_j  H_d   \nonumber \\ 
& + \Y^{\nu}_{ij} {\nu}^c_i   L_j  H_u  + \mu   H_u H_d   - \frac{1}{2} {M_{R}}_{i}{\nu}^c_i {\nu}^c_i ,
\end{align} 
where $i,j = \{1,2,3\} $ are generation indices. Note that the right handed neutrino Majorana mass matrix is diagonal. The SU(2) and SU(3) contractions are suppressed in the above Lagrangian.  
In the program,  the seesaw scale is taken to be $q^2 = M_{R_3}^2$.  At this scale, the right handed neutrino mass matrix can be diagonalized by a rotation of the right handed neutrino fields. The Dirac Yukawa matrix $\Y^\nu$ is defined at this scale in the basis where the right handed neutrino mass matrix is diagonal. \SUSEFLAV\ considers the inputs at the scale $M_{R_3}$ where right handed neutrino masses as well as the neutrino Dirac Yukawa coupling matrix. Finally it should be noted that in the present version of \SUSEFLAV, the ordering of the right handed neutrino mass eigenvalues is taken as  $M_{R_1} \lesssim M_{R_2} \lesssim M_{R_3}$. We have not included the option of inverted hierarchy for the right handed neutrinos in the present version. The program will abort if such a choice is made.  

Below the seesaw scale, once the right handed neutrinos are integrated out, we are left with the five dimensional operator  defining the the light neutrino mass
matrix as 
\begin{align}
\mathcal{W}\ =\ &
\Y^d_{ij} d^c_i Q_j  H_d +  \Y^u_{ij} u^c_i Q_j  H_u +
\Y^e_{ij} e^c_i  L_j  H_d   \nonumber \\ 
&+   \mu   H_u H_d   + \frac{\kappa_{ij}}{\Lambda}\;  L_i H_{u} L_j H_{u} 
\end{align} 
The light neutrino mass matrix is given  at the weak scale after the ${\rm SU(2)}_L \times {\rm U(1)}_Y$ breaking as
\begin{equation}
\mathcal{M}_\nu = \frac{\kappa_{ij}}{\Lambda}\; \langle H_{u}^{0} \rangle^{2} 
\end{equation} 
where $\Lambda$ represents the right handed neutrino mass scale or the seesaw scale and $\langle H_{u}^{0} \rangle$ is the vev of the $H_{u}$ superfield. The five dimensional operator is renormalized  from the seesaw 
scale to the electroweak scale. These corrections can be significant for inverse hierarchal and degenerate spectrum for neutrino masses \cite{antusch}. 
The present version of the program does not contain the renormalization for the light neutrinos, we refer users to use the existing programs like REAP \cite{antusch}. The main reason for not including the RG effects for light neutrinos has been the famous ambiguity in relating the light neutrino masses
to the neutrino Yukawa couplings \cite{casasibarra}. Furthermore, we are more interested in the effects of seesaw mechanism on the soft supersymmetric 
masses and couplings.  However, we do provide the option for the users to define the Yukawa matrices in terms of the Casas-Ibarra ${\cal R}$-parameterization \cite{casasibarra}. In this parametrization the neutrino Yukawa matrix $\Y^{\nu}$ is defined in terms of light neutrino masses as well as the right handed neutrino masses. In ${\cal R}$-parametrization, $\Y^{\nu}$ at the seesaw scale defined as
\begin{equation}
\Y^{\nu} = \frac{1}{\langle H_{u}^{0} \rangle^{2}}\ {\cal D}_{\sqrt{M_{R}}}\, {\cal R} \, {\cal D}_{\sqrt{\cal{M_{\nu}}}}\, {\bf U}^{\dagger}_{PMNS}
\end{equation}
where ${\bf U}_{{\rm PMNS}}$ is the Pontecorvo-Maki-Nakagawa-Sakata matrix \cite{pmns} and ${\cal R}$ is any $3 \times 3$ orthogonal matrix. The ${\cal D}_{\sqrt{M_{R}}}$ and ${\cal D}_{\sqrt{\cal{M_{\nu}}}}$ are defined as
\begin{align}
{\cal D}_{\sqrt{M_{R}}}\ =& \ {\rm diagonal}\left(\sqrt{M_{R_{1}}},\,\sqrt{M_{R_{2}}},\,\sqrt{M_{R_{3}}}\right) \\
{\cal D}_{\sqrt{\cal{M_{\nu}}}}\ =& \ {\rm diagonal}\left(\sqrt{\kappa_{1}},\, \sqrt{\kappa_{2}},\, \sqrt{\kappa_{3}}\right) 
\end{align}
where $\kappa_{1}$, $\kappa_{2}$ and $\kappa_{3}$ are light neutrino mass eigenvalues. It is important to note that $\cal R$ can be complex in nature also, but in the present version of \SUSEFLAV\ we take $\cal R$ to be real orthogonal matrix. One can parametrize the $\cal R$ matrix in terms of 3 angles but in \SUSEFLAV\ we have not parametrized the $\cal R$ matrix and left all the 9 elements of $\cal R$ as user defined input. Various other parametrization of the $\Y_{\nu}$ matrix have also been carried out in \cite{diffrparam}. Users interested in using the R-parameterization  can use \texttt{sinputs-rpar.in} in the main directory 
for their computations. 


In addition to the user defined $\Y^\nu$ at the seesaw scale, we also provide two other choices for $\Y^\nu$ and $M_{R}$ based on Grand Unified Models like SO(10). 
Both these cases consider hierarchal masses for the light neutrinos (${\mathcal M}_\nu$). These are
\begin{enumerate}

\item CKM Case:

In this case the $\Y^\nu$ and $M_{R}$ are given as 
\begin{align}
\Y^\nu =& \begin{pmatrix}
h_u &0 &0 \\
0 & h_c &0 \\
0&0& h_t \end{pmatrix}  {\bf V}_{\text{CKM}} \\
\text{Diagonal}[M_{R}] =& \{M_{R_3}, M_{R_2}, M_{R_1}\}= \{10^{14}, 10^9, 10^6\}~ \text{GeV} \nonumber
\end{align}
where $h_u,h_c,h_t$ are the Yukawa couplings of the up, charm and the top quarks and ${\bf V}_{\text{CKM}}$ is the quark sector mixing matrix.

\item PMNS Case:

In this case the $\Y^\nu$ and $M_{R}$ are given as 
\begin{eqnarray}
\Y^\nu &=& \left( \begin{array}{ccc}
h_u &0 &0 \\
0 & h_c &0 \\
0&0& h_t \end{array} \right) {\bf U}_{\text{PMNS}} \\
\text{Diagonal}[M_{R}] &=& \{M_{R_3}, M_{R_2}, M_{R_1}\}= \{10^{14}, 10^9, 10^6\}~ \text{GeV} \nonumber
\end{eqnarray}
where ${\bf U}_{{\rm PMNS}}$ is the leptonic mixing matrix. Both ${\bf V}_{\text{CKM}}$ and ${\bf U}_{\text{PMNS}}$ are defined in the \texttt{SuSemain.f} file in 
the {\tt src/} directory.  
Below the seesaw scale, $q^2 \lesssim M_{R_3}^2$, the right handed neutrinos (and sneutrinos) decouple from the theory, and the model is defined by MSSM.
\end{enumerate}

The soft supersymmetric breaking terms in the seesaw enhanced MSSM are same as in the MSSM together with the additional terms 
involving the right handed sneutrinos. These include the mass terms for the gauginos, mass squared terms for all the scalar particles 
and also the bilinear terms and trilinear terms:  
\begin{align}
\label{softlag}
-\mathcal{L}_{\text{soft}} \ \supset \ & \frac{1}{2} \left( M_1 \widetilde{B} \widetilde{B} + M_2 \widetilde{W} \widetilde{W} + M_3 \tilde{g} \tilde{g} \right) \nonumber \\
&+ m_{H_u}^2 |H_u|^2 + m_{H_d}^2 |H_d|^2 + \m_{\tilde{L}_{ij}}^2 \tilde{L}_i^\star \tilde{L}_j + \m^2_{\widetilde{e}^c_{ij}} \widetilde{e^{c}}^{\star}_i
\widetilde{e}^c_j + \m^2_{\widetilde{\nu^c}_{ij}} \widetilde{\nu^c}^\star_i \widetilde{\nu^c}_j + \ldots \nonumber  \\
&+ B_\mu H_u H_d +  {\bf B}_{M_{ij}} \widetilde{\nu^c_i} \widetilde{\nu^c_j} + h.c.   \nonumber \\
&+ \tilde{\A}_{ij}^u \widetilde{Q}_i \widetilde{u}^c_j H_u  + \tilde{\A}_{ij}^d \widetilde{Q}_i \widetilde{d}^c_j H_d + \tilde{\A}_{ij}^e \widetilde{L}_i \widetilde{e}^c_j
H_d + \tilde{\A}_{ij}^\nu \widetilde{L}_i \widetilde{\nu}^c_j H_u  
\end{align} 
We use the  factorization   $\tilde{\A}^u \equiv \A^u \Y^u$, for all the trilinear couplings at the weak scale\footnote{The RGEs are however, defined in terms of
 $\tilde \A^{f}$, which sets the format for inputs at the high scale. In mSUGRA, we have $\tilde \A^{f} = A_0 \textbf{1} $}. As noted before,  at the energies 
 $q^2 \lesssim M_{R_3}^2$, the right handed sneutrinos also decouple from the theory,
 along with the right handed neutrinos. In the program, we decouple the right handed neutrinos sequentially at different scales : the heaviest right handed neutrino 
 at $M_{R_3}$ and the second heaviest one at $M_{R_2}$ and the lightest one at$M_{R_1}$. 

The right handed neutrinos can be easily removed from the model to recover MSSM without right handed neutrinos, by choosing $\Y^\nu = \textbf{0}$. This automatically decouples the right handed neutrinos in the theory. An explicit  option is also provided in the input files, where by turning on/off the parameter {\tt rhn}, one can either include/remove right handed neutrinos in the theory.  
Finally, let us note that quantum effects above the scale of seesaw $q^2 \gg M_{R_3}^2$ will make the mass matrix $M_R$ non-diagonal. 
The running of the Majorana mass matrix can also effect the ${\bf B}_{M}$ term in the soft potential, described below, at the 1-loop level. 
The ${\bf B}_{M}$ can have implications for flavor physics and EDMs, if it is large through finite terms. These effects \cite{farzan} are not 
computed in \SUSEFLAV. 

The complete set of 2-loop RGE for a general superpotential and MSSM are presented in \cite{Martin:1993zk}. 
For the supersymmetric seesaw model, we use the RGE's from \cite{simonetto}. 


\section{SUSY Breaking Mechanisms}
\label{susybreak}
Supersymmetry is broken spontaneously in a hidden sector and is then communicated to the MSSM sector through the `messenger sector'. The messengers could be gauge interactions or  gravitational interactions. The result of this communication leads to soft supersymmetry breaking terms in the MSSM.  While the form of the soft Lagrangian Eq. (\ref{softlag}) is itself not dependent on the mediation mechanism, the physical quantities i.e., the masses, the couplings etc., are determined in terms of few `fundamental' parameters depending on the mediation mechanism. Popular among such supersymmetry breaking schemes are (i) minimal Supergravity (mSUGRA) (ii) Gauge Mediated Supersymmetry Breaking (GMSB) (iii) Anomaly Mediated Supersymmetry Breaking (AMSB) (iv) Gaugino mediation (v) Moduli mediation etc. among a host of other possibilities \cite{lutylectures}. In addition there could be variations within each of the above schemes. In the \SUSEFLAV, we have in-built (i) mSUGRA  and some  of its variations: (a) Non-Universal Higgs Models (NUHM), (b) Non-Universal Gaugino Models (NUGM) and (c) Complete Non-Universal Model (CNUM) and (ii) Gauge Mediated Supersymmetry Breaking (GMSB) models. The corresponding input files are  given in the {\tt examples/} directory. An input file with completely non-universal soft parameters is also presented for supergravity mediation, where the users can define the boundary conditions of their choice. For other supersymmetric breaking models, the users can modify appropriately the {\tt slha.in} file and run the program accordingly. In the section below we describe the two supersymmetric breaking scenarios considered in \SUSEFLAV.

\subsection{mSUGRA and its variations} 
Supersymmetry is broken spontaneously in a hidden sector and the communicated to the visible sector through the gravitational interactions. If the supergravity K\"ahler metric is canonical in matter fields, the soft terms resultant after integrating out the supergravity multiplet (while keeping the gravitino mass fixed),  are universal in nature \cite{nilles}. The property \textit{universal} refers to the flavor space \textit{i.e.} all the soft terms take the same value irrespective of the flavor at the mediation scale. In \SUSEFLAV\ we have considered the mediation scale to be $M_{\text{GUT}}$. At this scale, all the soft terms are determined by four parameters and the sign of the $\mu$-parameter.
\begin{enumerate}
\item
At $M_{\text{GUT}}$ the gaugino masses are universal to a value $M_{1/2}$, i.e.
\begin{equation}
M_{1}(M_{\text{GUT}}) =  M_{2}(M_{\text{GUT}}) = M_{3}(M_{\text{GUT}}) \equiv M_{1/2}
\end{equation}
\item
The scalar and the Higgs masses are given by the parameter $m_0^2$ at $M_{\text{GUT}}$.  
\begin{align}
\M^{2}_{{\tilde Q}_{i}} (M_{GUT})	 = \M^{2}_{{\tilde u}_{R_{i}}} (M_{GUT}) &= \M^{2}_{{\tilde d}_{R_{i}}} (M_{GUT}) = \M^{2}_{{\tilde L}_{i}} (M_{GUT}) = \M^{2}_{{\tilde l}_{i}}(M_{GUT}) \equiv m^{2}_{0}\; {\bf 1} \nonumber \\  m^{2}_{{H}_{u}} (M_{GUT}) &= m^{2}_{{H}_{d}} (M_{GUT}) \equiv m^{2}_{0}\;
\end{align}
\item
The trilinear couplings are given by the parameter  $A_{0}$  at $M_{\text{GUT}}$ 
\begin{equation}
{\tilde \A}^{u}_{ij}(M_{GUT}) = {\tilde \A}^{d}_{ij}(M_{GUT}) = {\tilde \A}^{l}_{ij}(M_{GUT}) \equiv A_{0} \; {\bf 1}_{ij}
\end{equation}
\end{enumerate}
To specify the spectrum at the weak scale, two more parameters need to be fixed. First is the ratio of the vacuum expectation values (vevs) of the
two Higgs fields,  $\tan\beta = v_{u}/v_{d}$. Second is a discrete parameter, the sign of $\mu$ or the Higgsino mass parameter. The magnitude
of $\mu$ is fixed by the radiative electroweak symmetry breaking mechanism which has been incorporated in the program. 

\subsubsection{Non-Universal Models}
In models based on Grand Unified theories, it has been proposed that  the strictly universal feature of the soft masses might not be valid and in fact some amount of non-universalities can enter in a model-dependent fashion. For example, in models where the hidden sector gauge kinetic function is no longer singlet under the GUT group, gaugino masses would become non-universal at the high scale. However the non-universalities enter in a predictive fashion when a particular gauge group is chosen, as the ratios of the gaugino masses are now fixed by the Clebsch-Gordan  coefficients of respective decomposition. These ratios are well known for the GUT models based on various gauge groups, e.g., SU(5) \cite{Ellis:1984bm,Ellis:1985jn,Drees:1985bx,Corsetti:2000yq}, SO(10) \cite{Chamoun:2001in,Bhattacharya:2009wv,Martin:2009ad}. 
Without resorting to any particular model,  we have incorporated non-universal gaugino mass scenario in \SUSEFLAV, by considering 
\begin{equation}
M_{1}(M_{GUT}) \;\neq\; M_{2}(M_{GUT}) \;\neq\; M_{3}(M_{GUT}).
\end{equation}
The user has the freedom of choosing any ratios among these three parameters at GUT scale. The corresponding input file is \texttt{sinputs-nugm.in}.
 A second class of non-universality  which has been incorporated in \SUSEFLAV\ is for the Higgs \cite{Ellis:2002iu}. It has been argued that since in Grand Unified theories (GUTs) like SO(10), all the matter sits in a single representation where as the Higgs sits in separate representation, the universality of the soft masses need not include Higgs, especially when supersymmetry breaking mediation happens close to the GUT scale.  It has also been realized that introducing such non-universality makes the $\mu$ parameter free and thus leading to completely different phenomenology at the weak scale especially for dark matter. Thus the boundary conditions at the high scale in our notation are given by 
\begin{align}
\M^{2}_{{\tilde Q}_{i}} (M_{GUT}) =\; & \M^{2}_{{\tilde u}_{R_{i}}} (M_{GUT}) =\;  \M^{2}_{{\tilde d}_{R_{i}}} (M_{GUT}) =\;  \M^{2}_{{\tilde L}_{i}} (M_{GUT}) = \M^{2}_{{\tilde l}_{i}}(M_{GUT}) \equiv m^{2}_{0}\; {\bf 1} \nonumber \\  
  & m^{2}_{{H}_{u}} (M_{GUT}) \equiv m^{2}_{10} \;\;\; ; \;\;\; m^{2}_{{H}_{d}} (M_{GUT}) \equiv m^{2}_{20}
\end{align}
Note that we intermittently use the  notation $m_{10}$ and $m_{20}$ for the Higgs mass parameters as defined above in the non-universal Higgs mass model. The input file for this case \texttt{sinputs-nuhm.in}. It should be noted that a  negative sign for any soft-mass as input would mean a negative sign for that soft-mass squared in the program. 

In addition to these non-universal input files, a completely generic input file called \texttt{sinputs-cnum.in} is provided where the user can provide \textit{all} the supersymmetric breaking parameters at the high scale (mSUGRA) without any assumptions on their structure. This input file is more suited for models with flavor structure at the high scale.

\subsection{GMSB}
The second class of supersymmetric breaking models incorporated in \SUSEFLAV\ is Gauge Mediated Supersymmetric Breaking (GMSB). As before, supersymmetry is broken spontaneously in the hidden sector, but now communicated to the MSSM sector through gauge interactions. The minimal set of models under this category goes under the name, Minimal Messenger Model (MMM) \cite{MMM}. In this model, a set of messenger superfields transforming as complete representations of SU(5) ($\supset \mathcal{G}_{\text{SM}}$) gauge group and couple directly to a singlet field which parameterized the supersymmetry breaking in the hidden sector. Supersymmetry breaking is then transmitted to the MSSM through SM gauge interactions.  The following superpotential represents the messenger sector coupling to the hidden sector  
\begin{equation}
{\cal W} = \lambda X \Phi_i \bar{\Phi}_i
\end{equation}
where $\Phi$ and $\bar{\Phi}$ are messenger sector superfields transforming as $5$ and $\bar{5}$ of SU(5) and $i$ runs for the number of messenger sector superfields, typically $i=[1,5]$. The $X$ superfield representing the hidden sector is parameterized by the vacuum expectation values for both its scalar component $\langle X\rangle$ as well as its auxiliary component $\langle F_X\rangle$. Gauge interactions with the messenger fields lead to gauginos attaining masses at 1-loop which at the Messenger scale are given by 
\begin{equation}
M_{a}\left(M_{mess}\right) = \frac{\alpha_{a}\left(M_{mess}\right)}{4\pi}\; \Lambda \; g\left(\frac{\Lambda}{M_{mess}}\right) \sum_{i}n_{a}(i);
\end{equation}
where the Messenger scale $M_{mess} = \lambda \langle X \rangle$ and $\Lambda = \langle F_{X} \rangle/ \langle X \rangle$. 
Here $n_{a}(i)$ is the Dynkin index for the messenger pair $\Phi, \overline{\Phi}$ and the sum runs over all the messengers in each group. 

The scalars attain their masses from the 2-loop diagrams. These are given as
\begin{equation}
m^{2}_{s}(M_{mess}) = 2 \Lambda^{2}\; f\left(\frac{\Lambda}{M_{mess}}\right) \; \sum_{a, i} n_{a}(i)\; C_{a} \; \left(\frac{\alpha_{a}(M_{mess})}{4\pi}\right)^{2} 
\end{equation}
Where $C_{a}$ is the quadratic Casimir invariant of the MSSM fields and the function $g(x)$ and $f(x)$ are defined in \cite{Martin:1996zb} and \cite{Dimopoulos:1996gy} respectively.
The leading order contribution to the tri-linear couplings comes from the 2-loop diagrams but they are suppressed by an extra $\alpha/4\pi$ factor compared to the gaugino masses. Thus to an very good approximation we can take 
\begin{equation}
{\tilde \A}^{u,d,l}_{ij} (M_{Mess}) \simeq 0
\end{equation}
Thus in minimal GMSB model considered in \SUSEFLAV\ we have  the following 5 parameters as inputs
\begin{equation}
\tan\beta,\; {\rm sign}(\mu),\; M_{mess},\; \Lambda {\rm ~and~} n.
\end{equation}
The input file \texttt{sinput-gmsb.in} and \texttt{mgmsb1.1.in} and so on, are built-in files to specify the inputs of the GMSB model in the {\tt examples/} directory
of \SUSEFLAV.


\section{Calculation of Supersymmetric Spectrum}
\label{algorithm}

\begin{figure}[htb]
\begin{center}
 \includegraphics[width=0.90\textwidth,angle=0]{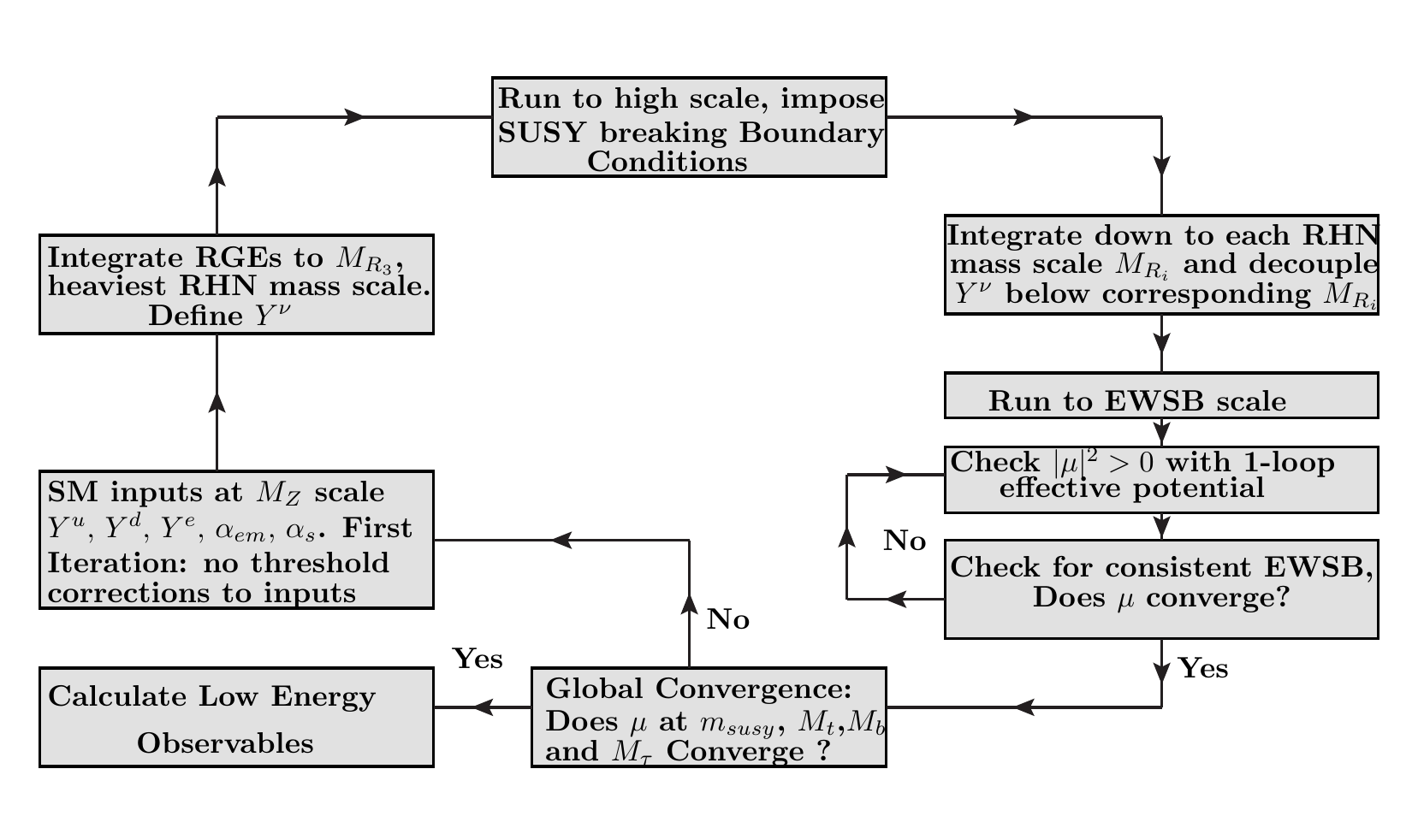}
\caption{Flowchart of the program }
\label{flowchart1}
\end{center}
\end{figure}

Once the user chooses  the particular model of supersymmetry breaking by typing in the various parameters in the relevant input file, the program computes the spectrum at the weak scale, checks for the various direct and indirect search limits and computes the various observables including the flavor violating ones like $\mu \to e + \gamma$. The computation of the spectrum involves several complicated intermediate steps which has been already explained in detail by various existing programs \cite{Allanach:2001kg,Djouadi:2002ze}. In \SUSEFLAV\ we follow a similar approach in computing the spectrum, however including flavor mixing as well as couplings with right handed neutrinos. In Fig. (\ref{flowchart1}), we have shown the flowchart of the computation of the spectrum and the observables in the program. We can summarize the computation in terms of three steps which however, are not  independent of each other as the procedure involves significant number of iterations.

\begin{itemize}
\item \textit{RGE evolution:} Using the MSSM RGE for the Yukawa and gauge couplings, run all the known SM parameters like gauge couplings, Yukawa couplings up to the scale  of supersymmetry breaking. In the case of mSUGRA, run to the scale where the gauge couplings corresponding to ${\rm SU}(2)_{L}$ ($g_{2}$) and ${\rm U}(1)_{Y}$ $\left(g_{1} \equiv \sqrt\frac{3}{5}\; g_{Y}\right)$ meet, this determines the GUT scale ($M_{{\rm GUT}}$).  At the SUSY breaking scale ($M_{\text{GUT}}$ in case of mSUGRA and $M_{\text{mess}}$  in case of GMSB), with the user defined input parameters and using the GUT scale SM parameters as the boundary conditions, run all the MSSM RGEs including those for the soft terms all the way down to $M_{{\rm SUSY}}$. For the initial run $M_{{\rm SUSY}}$ is defined to be 1 TeV. If seesaw mechanism is switched on, the program takes in to consideration the running of user defined neutrino Yukawa couplings between the seesaw scale and the supersymmetric breaking scale in both directions.

\item \textit{Radiative Electroweak symmetry breaking:} The resultant soft parameters at the $M_{\text{SUSY}}$ are used to check if they satisfy the tree level electroweak symmetry breaking conditions and compute the $\mu$ parameter. The full  one loop effective potential corrections are then computed  using the `tree level' $\mu$ parameter, which is then used to derive the 1-loop corrections to the $\mu$ parameter. This is repeated iteratively until the $\mu$ parameter converges. 

\item \textit{Convergence of the Spectrum:} In the final step, we run all the soft terms to the scale $M_Z$ where corrections to the SM parameters are added. We compute the supersymmetric corrections to the SM gauge couplings and the third generation fermion masses ($t$, $b$ and $\tau$). The resultant masses are fed in to the RGE routine as shown in Fig. (\ref{flowchart1}) and the soft spectrum is evaluated and run to MZ scale. 
This \textit{full} iteration is continued  until the SM third generation fermion masses converges to user defined precision (usually ${\cal O}(10^{-3}-10^{-4})$). Once this masses get converged we calculate various low energy observables (e.g. ${\rm BR}(\mu \to e \gamma)$, ${\rm BR}(b \to s \gamma)$ etc.). In section \ref{observables} we have discussed the low energy observables \SUSEFLAV\ calculates. 

\end{itemize}
In the following we describe each of these steps in more detail. 

\subsection{RGE Evolution} 
\begin{figure}[htb]
\begin{center}
\includegraphics[width=.90\textwidth,angle=0]{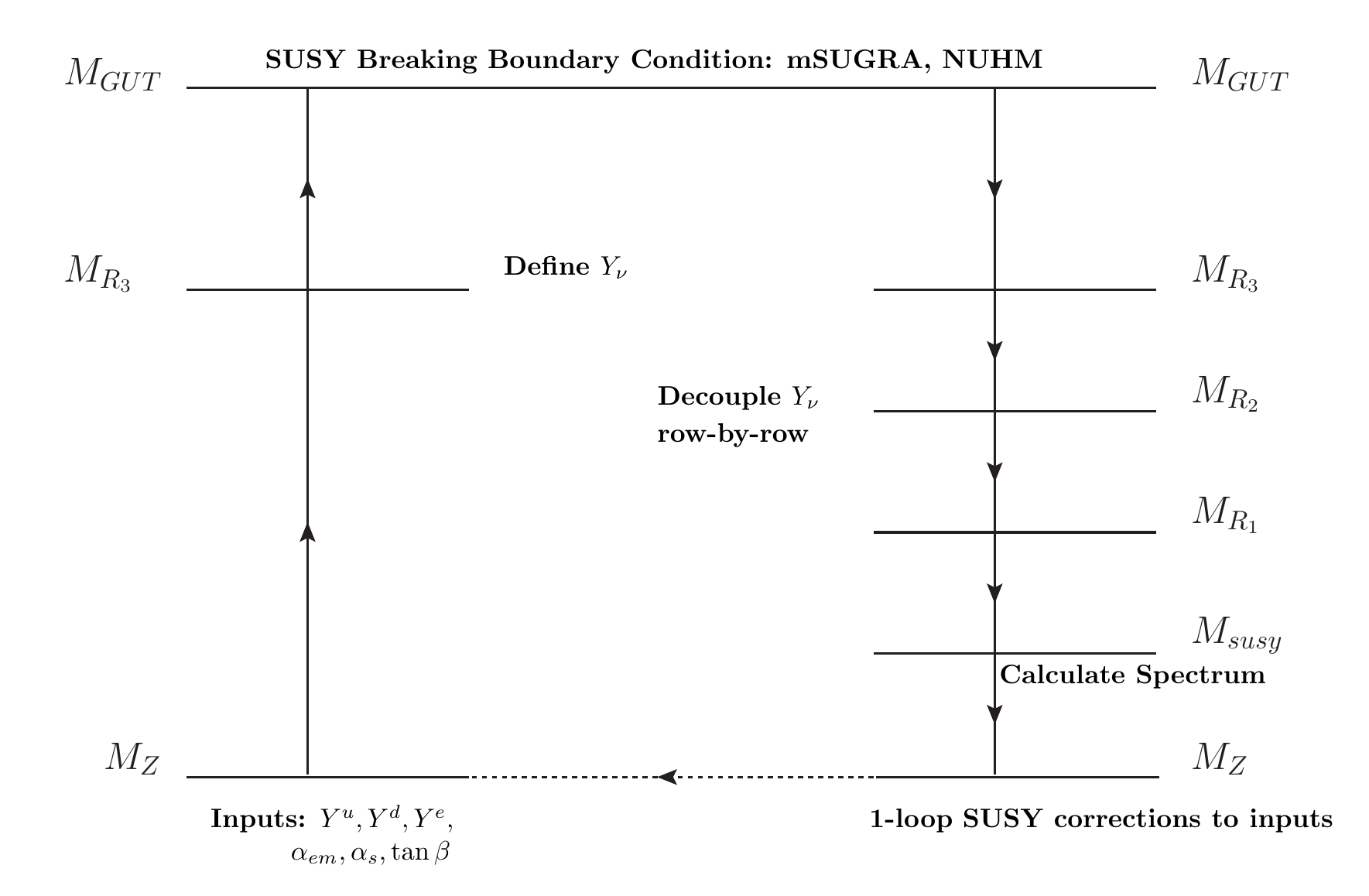}
\caption{Pictorial representation of evolution of RGEs}
\label{runrges}
\end{center}
\end{figure}

The standard model fermion masses and gauge couplings are the inputs to the program at the weak scale, taken to be equal to the Z-boson mass $M_Z$ in the program. The parameters are divided in to two subsets depending on whether radiative corrections are added or not. The parameters, masses of the first two generations quarks and leptons, for which we do not add radiative corrections are put in the file \texttt{src/stdinputs.h}. The values for these masses are taken from PDG 2012 \cite{pdg2010}. Most of the other parameters such as the leptonic mixing matrix, ${\bf U}_{\text{PMNS}}$, the hadronic mixing matrix (${\bf V}_{\text{CKM}}$)  are defined in the file  \texttt{src/SuSemain.f}. The $\overline{\text{MS}}$ values of Z-boson mass, the pole masses of the top quark $m_t^{pole}(m_{t})$ and tau lepton,  $m_{\tau}^{pole}(m_{\tau})$ and the $\overline{\text{MS}}$ mass of the bottom quark mass, $m_b^{\overline{\text{MS}}}(m_{b})$, are left as user defined inputs. The $\overline{\text{MS}}$ gauge couplings, electromagnetic, $\alpha_{em}(M_Z)$ and the strong coupling, $\alpha_s (M_Z)$ are also considered as inputs and are contained in the  input files. The $\overline{\text{MS}}$ inputs are converted to $\overline{\text{DR}}$ as the RGEs are written in the $\overline{\text{DR}}$ scheme. The conversion for the gauge couplings is given by 
\begin{equation}
\alpha_{em}^{\overline{\text{DR}}}(M_Z) = \left( \frac{1}{\alpha_{em}^{\overline{\text{MS}}} (M_Z)} - \frac{1}{6\pi} \right)^{-1},  \,\,\,
\alpha_{s}^{\overline{\text{DR}}}(M_Z) = \left( {\frac{1}{\alpha_{s}^{\overline{\text{MS}}}(M_Z)} - \frac{1}{4\pi}} \right)^{-1}
\end{equation}
The so defined $\alpha_{em}^{\overline{\text{DR}}}$ is in turn used to define the $\overline{\text{DR}}$ values of the $\alpha_{1,2}$. 
 \begin{equation}
\alpha_1(M_Z) \equiv \frac{g_{1}^{2}}{4\pi} = \frac{5 \alpha_{em}^{\overline{\text{DR}}}(M_Z)}{3 \cos^2 \theta_W} \;, \quad \alpha_2(M_Z) \equiv \frac{g_{2}^{2}}{4\pi} = \frac{\alpha_{em}^{\overline{\text{DR}}}(M_Z)}{\sin^2 \theta_W} 
\end{equation}
In a similar fashion, the bottom mass is converted from the $\overline{\text{MS}}$ to $\overline{\text{DR}}$ using 
\cite{kalmykov,Baer:2002ek} 
\begin{equation}
m_b^{\overline{\text{DR}}}(M_Z) = m_b^{\overline{\text{MS}}}(m_b)\cdot \left[1 -\frac{\alpha_s(M_Z)}{3\pi} - \frac{23\alpha_s^2(M_Z)}{72\pi^2} + \frac{3\alpha_2(M_Z)}{32\pi} + \frac{13\alpha_1(M_Z)}{288\pi}\right],
\label{mbmzdrbar}
\end{equation}
where the coupling constants appearing in the parenthesis are their $\overline{\text{DR}}$ values.  The masses of the tau lepton and top quark
are converted from their pole masses using the following relations:
\begin{eqnarray}
m_{\tau}^{\overline{\text{DR}}}(M_Z) &= & m_{\tau}^{pole}\cdot\left[1 - \frac{3}{8} \left(\alpha_1(M_Z) - \frac{\alpha_2(M_Z)}{4}\right)\right] \nonumber \\
m_t^{\overline{\text{DR}}}(M_Z)& =& m_t^{pole}\cdot \Delta_{m_t}^{QCD} 
\end{eqnarray}
where $\Delta_{m_t}^{QCD}$ is given by \cite{Bednyakov:2002sf} 
\begin{equation}
\Delta_{m_t}^{QCD}=1 - \frac{\alpha_s(m_t)}{3\pi}\, \left(5 - 3 \Delta_{tz}\right) - \alpha_s^{2}(m_t)\,\left(0.538 - \frac{43\Delta_{tz}}{24\pi} + \frac{3\Delta_{tz}^2}{8\pi^2}\right),
\label{mtdrbar}
\end{equation}
with $\Delta_{tz} = 2\ln \left(m_t^{pole}/M_Z \right)$ and 
\begin{equation}
\alpha_s(m_t) = \frac{\alpha_s^{\overline{\text{DR}}}(M_Z)}{1 + \frac{3}{4\pi} {\alpha_s^{\overline{\text{DR}}}(M_Z)\Delta_{tz}}}
\label{alphaemsdrbar}
\end{equation}
The $\overline{\text{DR}}$ corrected masses are used to define the $3 \times 3$ Yukawa matrices at the $M_Z$ scale which form the
inputs to the RGE. 
\begin{eqnarray}
\Y^u  = \frac{\sqrt{2}}{v  \sin \beta}\; \text{Diag}[m_u,m_c,m_t]\cdot {\bf V}_{CKM} \;  &;& \; \Y^d = \frac{\sqrt{2}}{v \cos \beta } \; \text{Diag}[m_d,m_s,m_b] \; ;  \nonumber\\
\Y^e = \frac{\sqrt{2}}{v \cos \beta} && \text{Diag}[m_e,m_\mu,m_\tau] 
\label{yu-coupling}	
\end{eqnarray}
We use the above defined Yukawas to run the full 2-loop RGEs from the weak scale ($M_Z$) up to the scale at which the two gauge couplings ($g_{1}$ and $g_{2}$) unify with an accuracy of 1\%. For GMSB scenario this scale is set by the user as messenger scale or $M_{\rm mess}$. In the case of right handed neutrinos three intermediate scales get introduced in the theory. As mentioned in the section \ref{mssmrn}, we consider the seesaw scale to be the mass of the heaviest right handed neutrino, $M_{R_3}$. At this scale we set the neutrino Yukawa ($\Y^{\nu}$) and run the RGEs with this Yukawa up to the high scale. At the high scale, depending on the model, we set the SUSY breaking boundary conditions and then run the RGEs down to the heaviest right handed neutrino mass scale i.e., $M_{R_{3}}$. Below this mass scale we decouple the heaviest right handed neutrino by setting its couplings in the neutrino Yukawa matrix  to zero. From $M_{R_{3}}$ we run down the RGEs to the next heaviest right handed neutrinos i.e., $M_{R_{2}}$ and then form $M_{R_{2}}$ to $M_{R_{1}}$. Below each of these scale, i.e., $M_{R_{2}}$ and $M_{R_{1}}$, we decouple the corresponding right handed neutrino by setting their couplings to zero. From $M_{R_{1}}$ we run the RGEs down to the scale  $M_{\rm SUSY}$. For the first iteration we take a guess value for $M_{\rm SUSY}$ which is 1 TeV. From the second iteration onwards the Electro-Weak Symmetry Breaking (EWSB) scale is set to the geometric mean of the two stop masses or $\sqrt{m_{\tilde{t}_1}\cdot m_{\tilde{t}_2}}$. At this scale we check for the EWSB condition and then we calculate the supersymmetric spectrum. A schematic picture of the integration of the RGE's in mSUGRA with seesaw mechanism is summarized  in Fig. (\ref{runrges}). In GMSB models, the high scale is the messenger scale, $M_{\rm mess}$ instead of $M_{\rm GUT}$ and the procedure of the integration is very similar. The seesaw mechanism can be incorporated in this class of models as long as heaviest right handed neutrino is lighter than the messenger scale ($M_{R_{3}} < M_{\rm mess}$).

\subsection{Radiative Electroweak Symmetry Breaking}
\label{ewsbchk}
The  tree level EWSB conditions at the $M_{\rm SUSY}$ scale are  defined as below
\begin{eqnarray}
\label{mu-ewsb}
\left|\mu\right|^2 &=&\frac{1}{2} \left[\tan 2\beta (m_{H_u}^2 \tan \beta - m_{H_d}^2\cot \beta) - M_Z^2\right] 
\nonumber \\
B_{\mu} &=& \frac{\sin 2\beta}{2} \left[m_{H_u}^2 + m_{H_d}^2 + 2|\mu|^2\right]
\end{eqnarray}
where $m_{H_u}^2$ and $m_{H_d}^2$ are the RGE output at $M_{\rm SUSY}$. For consistent electroweak symmetry breaking we require $\left|\mu\right|^2 >0$\footnote{Further, there should not be any tachyons in the physical Higgs spectrum.}. The tree level $|\mu|^{2}$ and RGE output of the other SUSY soft masses is used to calculate the tree level spectrum as described in \ref{app:tree}. Radiative corrections  can however significantly modify the tree level value of $\mu$.  Using tree level sparticle spectrum, we calculate the radiative corrections to the higgs potential up to one-loop order as given by BPMZ \cite{Pierce:1996zz}. The tadpoles modify the $m_{H_u}^2$ and $m_{H_d}^2$ as
\begin{equation}
m_{H_u}^2 \rightarrow m_{H_{u}}^2 - \frac{t_1}{v_1} \quad;\quad m_{H_d}^2 \rightarrow m_{H_d}^2 - \frac{t_2}{v_2}.
\end{equation}
With these radiatively corrected higges, using Eq. (\ref{mu-ewsb}), we calculate the radiatively corrected $|\mu|^{2}$. This is repeated iteratively
until  the convergence of $|\mu|^{2}$ reaches the  desired accuracy (default value is $\mathcal{O}(10^{-4})$). This accuracy level can be changed by changing the parameter \texttt{tol} in file \texttt{src/ewsbiterate.f}. There could be regions where the $\mu$ parameter does not converge within a small number of iterations. In such regions, the program considers the parameter point as $|\mu|$-non convergent. Once $|\mu|$ has converged we calculate $B_{\mu}$ as 
\begin{equation}
B_{\mu} = \frac{\sin 2\beta}{2} \left[\overline{m}_{H_u}^2 + \overline{m}_{H_d}^2 + 2|\mu|^2\right]
\end{equation}
It is important to note here that at $M_{\rm SUSY}$, \SUSEFLAV\ checks for D-flat directions in the potential as well as whether the potential is unbounded from below. It also checks for charge and color breaking minima. More details about these checks are discussed in the next section. 
Even if  these  conditions are not satisfied the program still proceeds to compute the spectrum  however, a flag is raised and written in the
output file. 

Once $\mu$ is converged, the program uses it to compute complete one loop corrections to the sparticle spectrum. We follow the work of BPMZ \cite{Pierce:1996zz} in computing these corrections\footnote{Current version of the program does not include flavor violating contributions from sleptons to all the 1-loop corrections. The sleptonic contributions are neglected in this version in the presence of flavor violation.}. Corrections to neutralinos and charginos are evaluated  at external momenta equal to $M_{\text{SUSY}}$ and the corrections to sfermion masses are calculated at an external momenta equal to their pole mass as prescribed by BPMZ. In \ref{oneloop} we have discussed more about these threshold corrections to the sparticles. As mentioned before all the parameters of the code are considered real, including the diagonalizing matrices. In determining  the neutral higgs masses the user has a choice to employ approximations for two loop which are mostly top-stop enhanced \cite{Heinemeyer:1999be} or full one loop tadpole  corrections as described in BPMZ or full one loop together with leading order two loop corrections. We have
implemented the two loop corrections due to  Slavich \textit{et. al} \cite{pietro} in our version 1.2.  This is the default choice for the Higgs spectrum in the present version. 
   
\subsection{Convergence of the Spectrum} 
In final step, the program evaluates the full one loop flavor conserving supersymmetric threshold corrections to SM parameters. The parameters which  are corrected are $m_t,\,m_b,\,m_{\tau},\,\alpha_s,\,\alpha_{em}$ and $\sin^2 \theta_{W}$.
One loop corrected running masses are given by the following,
\begin{equation}
\Delta_{m_t}(M_Z) = \Sigma_t^{BPMZ} + \Delta_{m_t}^{QCD},
\quad m_t(M_Z) = m_t^{pole}\left[1 + \Delta_{m_t}(M_Z)\right]
\label{oneloopmt}
\end{equation}

\begin{equation}
m_b(M_Z)^{\overline{DR}}_{MSSM} = \frac{m_b(M_Z)^{\overline{DR}}_{SM}}{1 + \Delta_{m_b}^{BPMZ}}
\label{oneloopmb}
\end{equation}

\begin{equation}
m_{\tau}(M_Z) = m_{\tau}^{pole}\left[1 + \Sigma_{\tau}^{BPMZ}\right]
\label{oneloopmtau}
\end{equation}
The quantity $\Sigma_t^{BPMZ}$, one loop correction to the top quark mass is evaluated at external momenta equal to  $m_t^{\overline{DR}}(M_{Z})$. Whereas,  $ \Delta_{m_b}^{BPMZ}$ and $\Sigma_{\tau}^{BPMZ}$ are evaluated in the limit of external momenta tending to zero. The expressions of these parameters are described in the \ref{oneloop}. The three gauge couplings get corrected as below 
\begin{equation}
\alpha_1(M_{Z}) = \frac{5 \alpha_{em}^{\overline{DR}}(M_{Z})}{3\left(1 - \Delta_{\alpha_{em}}\right) \cos^2 \theta_{W}} 
\label{oneloopa1}
\end{equation}
\begin{equation}
 \alpha_2(M_{Z}) = \frac{\alpha_{em}^{\overline{DR}}(M_{Z})}{\left(1 - \Delta_{\alpha_{em}}\right)\sin^2 \theta_{W}} 
\label{oneloopa2}
\end{equation}

\begin{equation}
 \alpha_3(M_{Z}) = \frac{\alpha_s(M_{Z})}{1 - \Delta_{\alpha_{s}}} 
\label{oneloopa3}
\end{equation}
where $\Delta_{\alpha_{em}}$ and $\Delta_{\alpha_{s}}$ are one loop corrections to the electromagnetic and strong coupling described in the \ref{oneloop}. Note that $\sin^2 \theta_W$ used in the above expressions is also radiatively corrected. Iterative method is implemented to correctly evaluate SUSY contributions to $\sin^{2} \theta_W$. The above corrected $\alpha_1(M_Z),\alpha_2(M_Z)$ and $\alpha_3(M_Z)$ and also the third generation SM fermions ($m_{t}$, $m_{b}$, $m_{\tau}$) masses are used as the input for the next long iteration. The iteration continues until the 
SM third generation fermions, namely top, bottom and tau mass are converged to user defined precision (usually ${\cal O}(10^{-3})$). This precision 
can be changed by the parameter named \texttt{spectrum tolerance} defined in the input files. Once the SM fermion masses get converged the program proceeds to
calculate the various low energy observables. Finally, lets note that if both $LL$ and $RR$ type leptonic flavor violation  is present, it could lead to corrections to lepton self energies \cite{nierste}, which
are not included in the present version of the code.


\section{Theoretical and Phenomenological Constraints}
\label{constraints}

The requirement of consistent evaluation of supersymmetric spectrum involves a check for theoretical constraints such as charge and color breaking minima (CCB), scalar potential unbounded from below (UFB) and efficient electroweak symmetry breaking at EWSB scale.

With every iterative step of the program \SUSEFLAV\ checks for CCB and UFB conditions at the tree level \cite{Frere:1983ag}. These conditions are governed by equations \ref{ccbcri}, \ref{ufbcri} and are simultaneously implemented while computing the tree level $\mu$ parameter (checking for efficient EWSB, see section \ref{ewsbchk} for the complete description).

\begin{align}
&{\rm {\bf CCB:}} \;\;\;\;  3 \left( m_{Q_{u}}^{2} + m_{f}^{2} + |\mu|^{2} + m_{H_{f}}^{2}\right) \geq \left|A_{f}\right|^{2}  \label{ccbcri}\\
&{\rm {\bf UFB:}}\;\;\;\;\;m_{H_{u}}^{2} + m_{H_{d}}^{2} + 2 |\mu|^{2} \geq 2 \left|B_{\mu}\right|^{2} {\;\rm at\;} Q>M_{EWSB}\label{ufbcri}
\end{align}

It is important to note that the complete MSSM spectrum is still calculated even if CCB and UFB conditions are not satisfied. However, if the supersymmetric scalar potential has a charge and color breaking minima which is lower than electroweak minimum a warning flag {\tt CCB} is generated. Similarly, if the complete supersymmetric scalar is unbounded from below a warning flag {\tt UFB} is generated. 

Apart form the above described theoretical constraints the program also imposes additional phenomenological constraints on the obtained spectrum. From the direct non-observation of charged dark matter in the Universe we require $m_{\tilde{\tau}}>m_{\chi^0}$ or LSP being neutral. Regions for which this condition is not true is excluded as $\tilde{\tau}$ LSP regions. Consequently the flag is marked as {\tt LSPSTAU}. Also, we require the the spectrum to be non-tachyonic. If tachyonic spectrum is encountered it is flagged as {\tt TACSPEC}. The program also indicates the sector  where the  tachyon occurs, for example: a tachyon in sleptonic sector is marked as {\tt TACSLP}.  Some  lower bounds on various sparticle masses that result from direct searches at colliders, e.g., the lightest Higgs mass $m_h>114.5 {\rm ~GeV}$, and the Chargino mass $m_{\chi^{\pm}} > 103.5 {\rm ~GeV}$ \cite{Barate:2003sz} are also incorporated in the program. Points failing to satisfy these bounds are flagged as {\tt LEPH} and {\tt LEPC} respectively. The present version does not include recent direct search limits from LHC \cite{LHC-leptonphoton}.

\section{Low Energy Observables}
\label{observables}

Once the complete supersymmetric spectrum is obtained we compute the following low energy observables in \SUSEFLAV.

\begin{itemize}
\item
{\bf Fine Tuning:} In MSSM the standard model masses and gauge couplings, {\it e.g.} $M_{Z}$, $m_{t}$, $m_{b}$ etc. are function of the input parameters of the model, i.e., for mSUGRA $m_{0},\ M_{1/2},\ A_{0},\ \tan\beta\ {\rm and}\ {\rm sign}(\mu)$. Now, we can recast the EWSB Eq. (\ref{mu-ewsb}) as below:

\begin{equation}
\label{finet}
M_{\bar{Z}}^{2} = -2|\mu|^{2} + \tan2\beta \left(m_{{H}_{u}}^{2}\tan\beta - m_{{H}_{d}}^{2}\cot\beta\right),
\end{equation}

where,  all the parameters on the right side of Eq. (\ref{finet}) are at the weak scale.  Given that $M_Z$ is know at a few percent level, it can be seen that  some amount of `tuning' between the parameters in the right side is needed. Various measures \cite{Barbieri:1987fn, de Carlos:1993yy,Anderson:1994dz, Agashe:1997kn}
have been proposed to quantify the `tuning' needed in some input parameter.
In \SUSEFLAV, we have followed Barbieri and Giudice \cite{Barbieri:1987fn, Agashe:1997kn}, in evaluating the  fine tuning for a given parameter $\lambda_{i}$ of the model as
\begin{equation}
\label{finet2}
\frac{\delta M_{Z}^{2}}{M_{Z}^{2}} = f(M_{Z}^{2}, \lambda_{i}) \; \frac{\delta \lambda_{i}}{\lambda_{i}}
\end{equation}
Where Barbieri-Giudice function or $f(M_{Z}^{2},\lambda_{i})$ is defined as
\begin{equation}
f(M_{Z}^{2}, \lambda_{i}) = \frac{\lambda_{i}}{M_{Z}^{2}} \; \frac{\partial M_{Z}^{2}}{\partial \lambda_{i}}
\end{equation}
We derive the fine tuning in $M_{Z}^{2}$ with respect to $\mu^{2}$ for which $f(M_{Z}^{2},\mu^{2})$ takes the following form
\begin{equation}
\label{ftmt}
f(M_{Z}^{2},\mu^{2}) = \frac{2|\mu|^{2}}{M_{Z}^{2}} \left[1 + \frac{(\tan^{2}\beta + 1)}{(\tan^{2}\beta -1)^{3}} \; \frac{4\tan^{2}\beta\;\left(m_{\overline{H}_{d}}^{2} - m_{\overline{H}_{u}}^{2}\right)}{\left(m_{\overline{H}_{d}}^{2} + m_{\overline{H}_{u}}^{2}\right)}\right]. 
\end{equation}
The fine tuning in $m_{t}$ with respect to $\mu^{2}$ or $f(M_{Z}^{2},m_{t})$ is expressed as follows
\begin{equation}
f(m_{t},\mu^{2}) =  \frac{1}{2} f(M_{Z}^{2},\mu^{2}) + \frac{1}{\tan^{2}\beta -1} \; \frac{2|\mu|^{2}}{m_{\overline{H}_{u}}^{2} + m_{\overline{H}_{d}}^{2}}
\end{equation}
\item
{\bf Electro-Weak Precision Measurement:} 
The electroweak observables $\rho$ parameter and $\sin^2\theta_W$ are defined in \ref{tree-rho} and \ref{tree-stw}.  
\begin{align}
\label{tree-rho}
\rho = \frac{M_W}{M_Z \cos \theta_{W}}\\ \label{tree-stw}
\sin^2 \theta_W = 1 - \frac{M^2_W}{M^2_Z}
\end{align}
At the tree level $\rho$ parameter is unity and $\sin^2\theta_W = 0.2286$. A deviation from these values are observed at one-loop level stemming from SM and supersymmetric corrections to $W$ and $Z$ boson masses. A measure of this deviation to $\rho$ parameter is given below 
\begin{align}
\label{ewprec}
\Delta \rho &= \frac{\Pi_{ZZ}(0)}{M_{Z}^{2}} - \frac{\Pi_{WW}(0)}{M_{W}^{2}} \\
{\rm and\;} \rho &= \frac{1}{(1-\Delta \rho)} 
\end{align}
The correction to $\sin^{2}\theta_{W}$ is described in \ref{oneloop}.
A precise measurement of these parameters by LEP, SLC and Tevetron imposes stringent constraints on $\Delta \rho$, requiring $\Delta \rho \lesssim 10^{-3}$ for a physically viable spectrum.

\item
{\bf BR($b\rightarrow s\gamma$) Constraint:} Another sector where the effect of SUSY particles can be seen is the radiative flavor changing decay of bottom quark, $b\rightarrow s\gamma$ \cite{Bertolini:1990if}. In the Standard model, this decay is mediated by loops containing up sector quarks and $W$ bosons. In supersymmetric theories, if CKM is the only source of hadronic flavor violation, additional contributions to $b\rightarrow s\gamma$ process come from chargino loops ($\widetilde{\chi}^{\pm}$), stop ($\tilde{t}$) squarks loops and charged Higgs bosons ($H^{\pm}$) loops. As in the perturbation theory the SM and supersymmetric correction appear at the same order, the measurement of BR($b\rightarrow s\gamma$) is a very powerful tool for constraining the SUSY parameter space. In \SUSEFLAV\ we have followed Bartl {\it et al.} \cite{Bartl:2001wc}, which includes Standard model NLO as well as MSSM LO contributions. Like the other observables, this branching ratio is calculated for all the valid points and written in the {\tt SLHA} file. More precise evaluations of this process are available publicly in the \cite{superiso, susyflavour,spheno2,bsgammacodes}.  However, given that we are concentrating on evaluation of flavor violating observables in the presence of flavor violation in the soft sector, we would like upgrade our computation of this processes in a future version to include gluino contributions which could become important
in the presence of squark flavor violating terms. Currently, we refer our users to couple the output of our program to one of the above existing codes depending on the
accuracy required and input parameters they are interested in. For example SuperIso \cite{superiso} computes the process at NNLO. It has an added advantage
that it can compute other B-physics observables also. As of now our program does not have an interface with SuperIso, but one can link to it using one
of its programs (\textbf{cmssm.c} etc ) to our code. 

\item
{\bf Anomalous Magnetic Moment of Muon $(g_{\mu}-2)$:} The muon anomalous magnetic moment ($g_{\mu}-2$) has been precisely measured by Muon $(g-2)$ collaboration \cite{Bennett:2002jb, Bennett:2004pv}. Due to supersymmetric particles in the loop, ($g_{\mu}-2$) gets non-negligible correction apart from the SM contribution. The experimentally measured value of anomalous magnetic moment of muon is:
\begin{equation}
a_{\mu}^{exp}\equiv \frac{(g_{\mu}-2)}{2} = (11659208 \pm 6) \times 10^{-10}. 
\end{equation}
Whereas, the difference in theoretical prediction by SM and experimental value, i.e $a_{\mu}^{exp} - a_{\mu}^{SM} = (28.7\pm 8.0)\times 10^{-10}$ \cite{Davier:2010nc}. 
Here the difference arises because of the fact of different estimates of hadronic vacuum polarization contribution. The contribution from the supersymmetric particles to $(g_{\mu}-2)$ is through $(\widetilde{\chi}^{0}-\tilde{\mu}/\tilde{\tau}/\tilde{e})$ loop or through $(\widetilde{\chi}^{\pm}-\tilde{\nu})$ loop. 
Supersymmetric parameter space can be severely constrained \cite{utpalc} as a supersymmetric explanation for the discrepancy with the experiment
prefers a `light'  supersymmetric spectrum. 
Moreover, both these contributions are $\tan\beta$ enhanced, so large values of $\tan\beta\,(\gtrsim30)$ are more severely constrained. 
We have taken the expression for one-loop contribution, due to supersymmetric particles, to the $(g_{\mu}-2)$  from Hisano {\it et al.} \cite{Hisano}. For a given set of input parameter \SUSEFLAV\ calculates the $(g_{\mu}-2)$ and writes it into the {\tt SLHA} file. Two loop contributions are not added in the present version which could be important in the very large tan$\beta$ regime \cite{dominik}.
One of the main reasons for not including these contributions is the absence of two loop computations including flavor violation in the sleptonic sector. The present version includes flavor 
violating contributions at the 1-loop level which become important when the seesaw is switched on. We hope to generalize the two loop contributions including flavor violations in a 
future version.

\item
{\bf Lepton Flavor Violating Decays:}
Subject to satisfying phenomenological constraints at $M_{\rm SUSY}$, we evaluate decay rates of rare lepton flavor violating processes.  The decay rates and branching ratios for the following processes are computed $\mu\,\rightarrow\, e\gamma$, $\tau\, \rightarrow\, e\gamma$, $\tau\, \rightarrow \, \mu \gamma$, $\mu\,\rightarrow\, e\gamma$, $\mu^{-}\,\rightarrow\, e^{+} e^{-} e^{-}$, $\tau^{-}\,\rightarrow\, \mu^{+} \mu^{-} \mu^{-}$, $\tau^{-}\, \rightarrow\, e^{+} e^{-} e^{-}$ and $\mu-e$ conversion rate in the nuclei.  
The search for charged lepton flavor violating decays can play a pivotal role in studying and discovering new physics beyond the standard model at TeV scale and above.
\begin{table}[htdp]
\caption{Present Experimental Bounds on LFV Processes}
\begin{center}
\renewcommand{\arraystretch}{1.5}
\begin{tabular}{|ccc|}
\hline \hline
Process & Experiment & Present upper bound \\ \hline
BR$(\mu\rightarrow e\, \gamma)$ & MEG \cite{meg11} & $2.4 \times 10^{-12} $\\
BR$(\mu\rightarrow e\, e\, e)$ & SINDRUM \cite{Bellgardt:1987du} &  $1.0 \times 10^{-12} $\\
CR$(\mu\rightarrow e\, {\rm in} \, {\bf Au})$ & SINDRUM-II \cite{Bertl:2006up} & $7 \times 10^{-13}$ \\
CR$(\mu\rightarrow e\, {\rm in} \, {\bf Ti})$ & SINDRUM-II \cite{Dohmen:1993mp} & $4.3 \times 10^{-12}$ \\
BR$(\tau\rightarrow \mu\, \gamma)$ & BABAR/Belle \cite{Asner:2010qj} & $4.4 \times 10^{-8}$ \\
BR$(\tau\rightarrow e\, \gamma)$ & BABAR/Belle \cite{Asner:2010qj} & $3.3 \times 10^{-8}$\\
BR$(\tau\rightarrow \mu\, \mu\, \mu)$ & BABAR/Belle \cite{Asner:2010qj} & $2.0 \times 10^{-8}$\\
BR$(\tau\rightarrow e\, e\, e)$ & BABAR/Belle \cite{Asner:2010qj} & $2.6 \times 10^{-8}$ \\ \hline \hline
\end{tabular}
\label{lfv-tab} 
\end{center}
\end{table}
The amplitudes of these processes are also sensitive to the left-right mixing in the sleptonic mass matrices in addition to the intergenerational mixing \cite{lee}. In calculating the amplitudes we closely follow the notations and expressions provided in \cite{Hisano, Hisano:1998fj}. The contribution to $\mu-e$ conversion rate in the nuclei is through penguin diagram where $\gamma$, $Z$ is exchanged and through box diagrams containing $\tilde{\chi}^0-\tilde{l}_i-q_{u,d}$ loops or $\tilde{\chi}^{-}-\tilde{\nu}_i-q_{u,d}$ loops. The current experimental upper bound on various LFV processes is tabulated in Table \ref{lfv-tab}.

In the current version of \SUSEFLAV\ we use the tree level masses and full $6\times6$ slepton mixing matrix to calculate LFV observables. The output is written in the {\tt BLOCK  SuSeFLAVLOWENERGY} in {\tt SLHA} format.

\item 
{\bf Light Neutral Higgs Boson Mass:} 
Recently, the ATLAS and CMS experiments at the Large Hadron Collider (LHC) at CERN have reported the discovery of a Higgs like
boson with a statistical significance of more than $5$ sigma. Both the experiments report  the mass of this particle to be close to 
125 GeV with the exact numbers given by : 
\begin{eqnarray}
&&m_h = 125.3 \pm 0.4 ~(\text{stat})  \pm 0.5~ (\text{sys}) ~~\text{GeV} ~~~~~~ \text{CMS \cite{cmshiggs}}~~ \nonumber \\
&&m_h = 126.0 \pm 0.4 ~(\text{stat})  \pm 0.4 ~(\text{sys}) ~~\text{GeV} ~~~~~~ \text{ATLAS \cite{atlashiggs}}~~ \nonumber 
\end{eqnarray}
In MSSM, this would correspond to the mass of lightest neutral  CP-even Higgs boson provided it has  SM-like couplings \cite{mssmhiggs1}.  
Both these requirements  put a strong constraints on most models of
supersymmetry breaking especially the ones with universal boundary conditions\cite{mssmhiggs2}. In the universal class of models,
large values of the trilinear couplings are required to generate a light higgs mass of that order. 
Two loop  contributions to the higgs mass matrices are important  and cannot be neglected while
studying the implications  on model parameter space.  As mentioned earlier,
\SUSEFLAV  computes the light higgs mass at the two loop level making it appropriate for analysis of this constraint
over a wide range of supersymmetry breaking models. In particular, implications on the supersymmetric
seesaw models and the complementarity with flavour processes  has already been studied \cite{our1} assuming universal
and non-universal higgs boundary conditions. As of now the light higgs mass constraint is not imposed automatically in the
program. Since there are still some theoretical and experimental uncertainties which are subject to modification as more
data pours in from LHC experiments, the exact range of the light higgs mass is left to the user. 

\end{itemize}
\section{Executing \SUSEFLAV}
\label{exec}
 Instruction to compile and install the package is provided in {\tt README} file.
 \SUSEFLAV\ uses the {\tt Diag} package by T. Hanh to diagonalize the mass matrices\footnote{Note that \LAPACK\ \cite{lapack} library is a another option for an earlier version of \SUSEFLAV\ for diagonalizing mass matrices. It is available on the website for downloading. We do not recommend the old version.}.
 \SUSEFLAV\ package produces three executables when compiled, namely {\tt suseflav}, {\tt suseflavslha} and {\tt suseflavscan} in the {\tt bin/} sub-directory. To compute the spectrum for a single point the usage of executables {\tt suseflav} and {\tt suseflavslha} is recommended. The user  must modify the corresponding input files {\tt sinputs.in} and {\tt slha.in} where all the input parameters are specified. To scan the parameter space the usage  of the executable {\tt suseflavscan} is recommended. The corresponding input parameters are in the input file {\tt sinputs\_scan.in}.
   
\subsection{{\bf Sample Input/Output}}
In this section we provide examples of input files which can be used to run the program. \SUSEFLAV\ has two input/output modes. This can be broadly classified into {\tt SLHA} I/O interface and non-{\tt SLHA} or traditional \SUSEFLAV\ I/O interface. The directory {\tt examples/} contains a variety of input files with which the user can run different models. 

\subsubsection{{\bf {\tt SLHA} Interface}} The main source file is {\tt runslha.f}.  To use the {\tt SLHA} interface the user should execute {\tt suseflavslha} by modifying the corresponding input file {\tt slha.in}.  Note that the user must rename the required {\tt SLHA} file as {\tt slha.in} to use that particular file as input. To run  the program type the following command at the terminal

{\tt ./suseflavslha}

\noindent
This sample input/output corresponds to point {\tt 10.1.5} \cite{AbdusSalam:2011fc} with Type-I see-saw mechanism with maximal mixing in mSUGRA/CMSSM scenario.
\begin{verbatim}
Block MODSEL                 # Select model
    1    1                   # sugra
    6    1                   # Quark Flavor Violation
Block SMINPUTS               # Standard Model inputs
    1   1.279340000E+02      # alpha^(-1) SM MSbar(MZ)
    2   1.166390000E-05      # G_Fermi
    3   1.172000000E-01      # alpha_s(MZ) SM MSbar
    4   9.118760000E+01      # MZ(pole)
    5   4.230000000E+00      # mb(mb) SM MSbar
    6   1.727000000E+02      # mtop(pole)
    7   1.770000000E+00      # mtau(pole)
Block MINPAR                 # Input parameters
    1   1.750000000E+02      # m0
    2   7.000000000E+02      # m12
    3   1.000000000E+01      # tanb
    4   1.000000000E+00      # sign(mu)
    5   0.000000000E+00      # A0
Block SUSEFLAV            # Algorithm specific inputs
 1   1.000000000E-03      # spectrum tolerance
 5   2.000000000E+00      # 2-loop running (1 = 1loop; 2 = 2loop)
 3   1.000000000E+00      # print Control (0 = do not print output; 1 = print output)
 2   1.000000000E+00      # Right handed neutrino (1 = yes; 0 = no)
 4   1.000000000E+00      # RHN mixing : 1 = ckm; 2 =PMNS; 3 = user defined
 7   1.000000000E+06      # MR1, Lightest rhn decoupling scale
 8   1.000000000E+09      # MR2, second lightest rhn decoupling scale
 9   1.000000000E+14      # MR3, Heaviest rhn decoupling scale
 10  0.000000000E+00      # Dirac neutrino mixing matrix 1,1
 11  0.000000000E+00      # Dirac neutrino mixing matrix 1,2
 12  0.000000000E+00      # Dirac neutrino mixing matrix 1,3
 13  0.000000000E+00      # Dirac neutrino mixing matrix 2,1
 14  0.000000000E+00      # Dirac neutrino mixing matrix 2,2
 15  0.000000000E+00      # Dirac neutrino mixing matrix 2,3
 16  0.000000000E+00      # Dirac neutrino mixing matrix 3,1
 17  0.000000000E+00      # Dirac neutrino mixing matrix 3,2
 18  0.000000000E+00      # Dirac neutrino mixing matrix 3,3
\end{verbatim}

\noindent The corresponding {\tt SLHA} output is generated in {\tt slha.out}
\begin{verbatim}
# Spectrum Output in SUSY Les Houches Accord 2
# SuSeFLAV v1.2.0
# D. Chowdhury, R. Garani and S. K. Vempati, hep-ph/1109.3551 
# For bug reports or any other queries please send email to suseflav@cts.iisc.ernet.in 
# Created on 12.06.2012 at 17:30 Hrs
#
BLOCK SPINFO  #  Program information
     1   SuSeFLAV     # Spectrum calculator
     2   1.2.0        # Version number
#
BLOCK MODSEL  #  MODEL NAME
     1      1   # mSUGRA
     3      4   # mSUGRA + RHN   
     6      3   # Lepton and quark flavor is violated
#
BLOCK MINPAR  # Input parameters
         1     1.75000000E+02   #  m0                 
         2     7.00000000E+02   #  m_1/2              
         3     1.00000000E+01   #  tanbeta(mz)        
         4     1.00000000E+00   #  sign(mu)           
         5     0.00000000E+00   #  A0                 
#
BLOCK SMINPUTS  # Standard Model inputs
         1     1.27934000E+02   # alpha_em (M_Z)^MSbar
         2     1.16639000E-05   # G_F [GeV^-2]
         3     1.17200000E-01   # alpha_S(M_Z)^MSbar
         4     9.11876000E+01   # M_Z pole mass
         5     4.23000000E+00   # mb(mb)^MSbar
         6     1.72700000E+02   # mt pole mass
         7     1.77000000E+00   # mtau pole mass
#
BLOCK EXTPAR  # Extra Input parameters
         0     1.97682880E+16   # Unification Scale
#
BLOCK MASS  # Mass Spectrum
# PDG code           mass       particle
        24     8.05095481E+01   # W+
        25     1.16435226E+02   # h
        35     9.79136838E+02   # H
        36     9.78810446E+02   # A
        37     9.82557324E+02   # H+
   1000001     1.29049183E+03   # ~d_1
   1000003     1.33317083E+03   # ~d_2
   1000005     1.33782720E+03   # ~d_3
   2000001     1.33783952E+03   # ~d_4
   2000003     1.39947428E+03   # ~d_5
   2000005     1.39948751E+03   # ~d_6
   1000002     1.08957331E+03   # ~u_1
   1000004     1.32515013E+03   # ~u_2
   1000006     1.34374200E+03   # ~u_3
   2000002     1.34410204E+03   # ~u_4
   2000004     1.39733996E+03   # ~u_5
   2000006     1.39738676E+03   # ~u_6
   1000011     3.12007210E+02   # ~l_1
   1000013     3.18856813E+02   # ~l_2
   1000015     3.18880311E+02   # ~l_3
   2000011     4.94924639E+02   # ~l_4
   2000013     4.96469065E+02   # ~l_5
   2000015     4.96470193E+02   # ~l_6
   1000012     4.87191917E+02   # ~nu_1
   1000014     4.90375557E+02   # ~nu_2
   1000016     4.90381909E+02   # ~nu_3
   1000021     1.57599127E+03   # ~g
   1000022     2.89580143E+02   # ~chi_10
   1000023     5.44321259E+02   # ~chi_20
   1000025     8.58605742E+02   # ~chi_30
   1000035     8.69437471E+02   # ~chi_40
   1000024     5.33851586E+02   # ~chi_1+
   1000037     8.62050764E+02   # ~chi_2+
#
BLOCK NMIX  # Neutralino Mixing Matrix
  1  1     9.97666897E-01   # N_11
  1  2    -1.03658855E-02   # N_12
  1  3     6.20427374E-02   # N_13
  1  4    -2.65332008E-02   # N_14
  2  1     2.32098394E-02   # N_21
  2  2     9.80197789E-01   # N_22
  2  3    -1.61409861E-01   # N_23
  2  4     1.12340793E-01   # N_24
  3  1     2.46412419E-02   # N_31
  3  2    -3.55274833E-02   # N_32
  3  3    -7.04937074E-01   # N_33
  3  4    -7.07950795E-01   # N_34
  4  1    -5.92863909E-02   # N_41
  4  2     1.94531848E-01   # N_42
  4  3     6.87867194E-01   # N_43
  4  4    -6.96764815E-01   # N_44
#
BLOCK UMIX  # Chargino Mixing Matrix U
  1  1     9.74989137E-01   # U_11
  1  2    -2.22252520E-01   # U_12
  2  1     2.22252520E-01   # U_21
  2  2     9.74989137E-01   # U_22
#
BLOCK VMIX  # Chargino Mixing Matrix V
  1  1     9.88224902E-01   # V_11
  1  2    -1.53008313E-01   # V_12
  2  1     1.53008313E-01   # V_21
  2  2     9.88224902E-01   # V_22
#
BLOCK USQMIX  # squark Mixing Matrix
  1  1     1.10996359E-03   # USQMIX_11
  1  2    -5.26653686E-03   # USQMIX_12
  1  3     3.55285163E-01   # USQMIX_13
  1  4     8.74704256E-04   # USQMIX_14
  1  5    -4.21826454E-03   # USQMIX_15
  1  6     9.34732542E-01   # USQMIX_16
  2  1     1.12595131E-02   # USQMIX_21
  2  2    -5.57113248E-02   # USQMIX_22
  2  3     9.23559656E-01   # USQMIX_23
  2  4     2.86784533E-02   # USQMIX_24
  2  5    -1.38092429E-01   # USQMIX_25
  2  6    -3.52015712E-01   # USQMIX_26
  3  1     2.97503603E-05   # USQMIX_31
  3  2    -9.98459110E-05   # USQMIX_32
  3  3     1.03612223E-03   # USQMIX_33
  3  4     9.77509497E-01   # USQMIX_34
  3  5     2.10888527E-01   # USQMIX_35
  3  6    -3.57456096E-04   # USQMIX_36
  4  1     2.18713035E-03   # USQMIX_41
  4  2     2.22241357E-03   # USQMIX_42
  4  3     1.33658497E-01   # USQMIX_43
  4  4    -2.08919254E-01   # USQMIX_44
  4  5     9.67647238E-01   # USQMIX_45
  4  6    -4.62304107E-02   # USQMIX_46
  5  1     9.82212602E-01   # USQMIX_51
  5  2     1.87753700E-01   # USQMIX_52
  5  3    -9.34002684E-04   # USQMIX_53
  5  4     5.08408180E-04   # USQMIX_54
  5  5    -2.40125319E-03   # USQMIX_55
  5  6     2.35206342E-04   # USQMIX_56
  6  1    -1.87418278E-01   # USQMIX_61
  6  2     9.80618229E-01   # USQMIX_62
  6  3     5.42538107E-02   # USQMIX_63
  6  4     2.10966003E-03   # USQMIX_64
  6  5    -9.57981337E-03   # USQMIX_65
  6  6    -1.49190697E-02   # USQMIX_66
#
BLOCK DSQMIX  # squark Mixing Matrix
  1  1     7.97027235E-03   # DSQMIX_11
  1  2    -3.83864371E-02   # DSQMIX_12
  1  3     9.76209913E-01   # DSQMIX_13
  1  4    -1.04373101E-05   # DSQMIX_14
  1  5    -2.41872348E-04   # DSQMIX_15
  1  6     2.13253615E-01   # DSQMIX_16
  2  1    -2.81490374E-03   # DSQMIX_21
  2  2     1.35643992E-02   # DSQMIX_22
  2  3    -2.12865830E-01   # DSQMIX_23
  2  4     1.79277418E-05   # DSQMIX_24
  2  5     9.89907963E-04   # DSQMIX_25
  2  6     9.76982723E-01   # DSQMIX_26
  3  1     6.10926291E-06   # DSQMIX_31
  3  2     5.89431513E-03   # DSQMIX_32
  3  3     6.86044622E-04   # DSQMIX_33
  3  4    -2.59132138E-05   # DSQMIX_34
  3  5     9.99981946E-01   # DSQMIX_35
  3  6    -9.45553891E-04   # DSQMIX_36
  4  1     2.96437790E-04   # DSQMIX_41
  4  2    -7.98684964E-07   # DSQMIX_42
  4  3     1.15232653E-05   # DSQMIX_43
  4  4     9.99999956E-01   # DSQMIX_44
  4  5     2.58944876E-05   # DSQMIX_45
  4  6    -1.50004569E-05   # DSQMIX_46
  5  1    -1.75678721E-01   # DSQMIX_51
  5  2     9.83552600E-01   # DSQMIX_52
  5  3     4.12376560E-02   # DSQMIX_53
  5  4     5.24615573E-05   # DSQMIX_54
  5  5    -5.82957991E-03   # DSQMIX_55
  5  6    -5.17098306E-03   # DSQMIX_56
  6  1     9.84411220E-01   # DSQMIX_61
  6  2     1.75875037E-01   # DSQMIX_62
  6  3    -1.15326259E-03   # DSQMIX_63
  6  4    -2.91633802E-04   # DSQMIX_64
  6  5    -1.04177570E-03   # DSQMIX_65
  6  6     1.44250146E-04   # DSQMIX_66
#
BLOCK SELMIX  # Slepton Mixing Matrix
  1  1     1.05172917E-05   # SELMIX_11
  1  2    -5.07243990E-05   # SELMIX_12
  1  3     1.03659288E-01   # SELMIX_13
  1  4     6.00795431E-08   # SELMIX_14
  1  5    -1.11136067E-05   # SELMIX_15
  1  6     9.94612864E-01   # SELMIX_16
  2  1    -3.08963419E-08   # SELMIX_21
  2  2     6.24290695E-03   # SELMIX_22
  2  3    -2.02732368E-06   # SELMIX_23
  2  4    -6.78386431E-09   # SELMIX_24
  2  5     9.99980513E-01   # SELMIX_25
  2  6     1.17032561E-05   # SELMIX_26
  3  1     2.98889602E-05   # SELMIX_31
  3  2    -6.33346635E-11   # SELMIX_32
  3  3     1.18627921E-08   # SELMIX_33
  3  4     1.00000000E+00   # SELMIX_34
  3  5     6.78606456E-09   # SELMIX_35
  3  6    -6.19572827E-08   # SELMIX_36
  4  1     9.84024127E-03   # SELMIX_41
  4  2    -4.73545901E-02   # SELMIX_42
  4  3     9.93448538E-01   # SELMIX_43
  4  4    -3.12319765E-07   # SELMIX_44
  4  5     2.98862230E-04   # SELMIX_45
  4  6    -1.03540457E-01   # SELMIX_46
  5  1     9.99950259E-01   # SELMIX_51
  5  2     2.09161645E-03   # SELMIX_52
  5  3    -9.70065714E-03   # SELMIX_53
  5  4    -2.98872962E-05   # SELMIX_54
  5  5    -1.30585027E-05   # SELMIX_55
  5  6     1.00054245E-03   # SELMIX_56
  6  1    -1.62739211E-03   # SELMIX_61
  6  2     9.98856442E-01   # SELMIX_62
  6  3     4.71237978E-02   # SELMIX_63
  6  4     4.78864799E-08   # SELMIX_64
  6  5    -6.23573698E-03   # SELMIX_65
  6  6    -4.86038870E-03   # SELMIX_66
#
BLOCK SNUMIX  # Sneutrino Mixing Matrix
  1  1     4.85609162E-03   # SNUMIX_11
  1  2    -2.34212181E-02   # SNUMIX_12
  1  3     9.99713892E-01   # SNUMIX_13
  2  1    -5.74301350E-02   # SNUMIX_21
  2  2     9.98069089E-01   # SNUMIX_22
  2  3     2.36616496E-02   # SNUMIX_23
  3  1     9.98337717E-01   # SNUMIX_31
  3  2     5.75286069E-02   # SNUMIX_32
  3  3    -3.50163123E-03   # SNUMIX_33
#
BLOCK ALPHA  # Higgs mixing
          -1.05473355E-01   # Mixing angle in the neutral Higgs boson sector
#
BLOCK HMIX Q=  1.20195104E+03  # DRbar Higgs Parameters
     1     8.50892351E+02   # mu(Q)
     2     9.63860969E+00   # tanbeta(Q)
     3     2.43818892E+02   # vev(Q)
     4     9.93075833E+05   # MA^2(Q)
#
BLOCK GAUGE Q=  1.20195104E+03  # The gauge couplings
     1     3.62957251E-01   # gprime(Q) DRbar
     2     6.42610014E-01   # g(Q) DRbar
     3     1.04880243E+00   # g_3(Q) DRbar
#
BLOCK TU Q=  1.20195104E+03  # The trilinear couplings
  1  1    -1.26688884E-01   # TURG_11
  1  2    -2.93067837E-02   # TURG_12
  1  3    -3.97143909E-04   # TURG_13
  2  1     2.93036897E+01   # TURG_21
  2  2    -1.26686864E+02   # TURG_22
  2  3    -5.37734251E+00   # TURG_23
  3  1    -1.11137936E+02   # TURG_31
  3  2     5.35228445E+02   # TURG_32
  3  3    -1.29451330E+04   # TURG_33
#
BLOCK TD Q=  1.20195104E+03  # The trilinear couplings
  1  1    -6.05492867E+00   # TDRG_11
  1  2    -1.02056677E-04   # TDRG_12
  1  3     2.38753240E-03   # TDRG_13
  2  1    -2.04113446E-03   # TDRG_21
  2  2    -1.21088577E+02   # TDRG_22
  2  3    -2.29965104E-01   # TDRG_23
  3  1     1.22702682E+00   # TDRG_31
  3  2    -5.90930668E+00   # TDRG_32
  3  3    -2.95665819E+03   # TDRG_33
#
BLOCK TE Q=  1.20195104E+03  # The trilinear couplings
  1  1    -1.44939639E-01   # TERG_11
  1  2    -2.31067898E-08   # TERG_12
  1  3    -4.08400006E-07   # TERG_13
  2  1    -4.82631701E-06   # TERG_21
  2  2    -3.02751406E+01   # TERG_22
  2  3     4.10638455E-04   # TERG_23
  3  1    -1.53360878E-03   # TERG_31
  3  2     7.38111171E-03   # TERG_32
  3  3    -5.16767721E+02   # TERG_33
#
BLOCK YU Q=  1.20195104E+03  # The top Yukawa coupling
  1  1     6.44039888E-06   # YU_11
  1  2     1.48968552E-06   # YU_12
  1  3     2.41182558E-08   # YU_13
  2  1    -1.48971566E-03   # YU_21
  2  2     6.44041987E-03   # YU_22
  2  3     2.72567759E-04   # YU_23
  3  1     7.22768067E-03   # YU_31
  3  2    -3.48078574E-02   # YU_32
  3  3     8.43790071E-01   # YU_33
#
BLOCK YD Q=  1.20195104E+03  # The down Yukawa coupling
  1  1     2.53400760E-04   # YD_11
  1  2    -1.10093360E-09   # YD_12
  1  3     2.56831364E-08   # YD_13
  2  1    -2.20186781E-08   # YD_21
  2  2     5.06812345E-03   # YD_22
  2  3    -2.47377375E-06   # YD_23
  3  1     1.31978305E-05   # YD_31
  3  2    -6.35600675E-05   # YD_32
  3  3     1.31856452E-01   # YD_33
#
BLOCK YE Q=  1.20195104E+03  # The tau Yukawa coupling
  1  1     2.77703519E-05   # YE_11
  1  2    -1.17975032E-11   # YE_12
  1  3     4.03483137E-10   # YE_13
  2  1    -2.46439380E-09   # YE_21
  2  2     5.80081939E-03   # YE_22
  2  3    -4.05872732E-07   # YE_23
  3  1     1.45732796E-06   # YE_31
  3  2    -7.01783388E-06   # YE_32
  3  3     9.96765043E-02   # YE_33
#
BLOCK MSOFT Q=  1.20195104E+03  # soft SUSY breaking masses at scale Q
         1     2.95504689E+02   # M_1
         2     5.45936065E+02   # M_2
         3     1.53262491E+03   # M_3
        21     2.14879878E+05   # M^2_Hd
        22    -6.96254426E+05   # M^2_Hu
        31     4.94404728E+02   # M_eL
        32     4.94396817E+02   # M_muL
        33     4.91242768E+02   # M_tauL
        34     3.15861187E+02   # M_eR
        35     3.15846380E+02   # M_muR
        36     3.11472426E+02   # M_tauR
        41     1.39829222E+03   # M_q1L
        42     1.39811849E+03   # M_q2L
        43     1.29136113E+03   # M_q3L
        44     1.34421791E+03   # M_uR
        45     1.34420302E+03   # M_cR
        46     1.11340022E+03   # M_tR
        47     1.33760077E+03   # M_dR
        48     1.33759064E+03   # M_sR
        49     1.33101944E+03   # M_bR
#
BLOCK MSQ2 Q=  1.20195104E+03  # M^2_Q soft SUSY breaking masses
  1  1     1.95522114E+06   # mSQRG_11
  1  2     1.03013845E+02   # mSQRG_12
  1  3    -2.38957910E+03   # mSQRG_13
  2  1     1.03013845E+02   # mSQRG_21
  2  2     1.95473530E+06   # mSQRG_22
  2  3     1.15080756E+04   # mSQRG_23
  3  1    -2.38957910E+03   # mSQRG_31
  3  2     1.15080756E+04   # mSQRG_32
  3  3     1.66761356E+06   # mSQRG_33
#
BLOCK MSU2 Q=  1.20195104E+03  # M^2_U soft SUSY breaking masses
  1  1     1.80692178E+06   # mSURG_11
  1  2    -5.84935042E-06   # mSURG_12
  1  3    -1.19615520E-02   # mSURG_13
  2  1    -5.84935042E-06   # mSURG_21
  2  2     1.80688177E+06   # mSURG_22
  2  3     4.19949204E+00   # mSURG_23
  3  1    -1.19615520E-02   # mSURG_31
  3  2     4.19949204E+00   # mSURG_32
  3  3     1.23966005E+06   # mSURG_33
#
BLOCK MSD2 Q=  1.20195104E+03  # M^2_D soft SUSY breaking masses
  1  1     1.78917582E+06   # mSDRG_11
  1  2    -9.47987536E-06   # mSDRG_12
  1  3     5.70618349E-03   # mSDRG_13
  2  1    -9.47987536E-06   # mSDRG_21
  2  2     1.78914873E+06   # mSDRG_22
  2  3    -5.49615942E-01   # mSDRG_23
  3  1     5.70618349E-03   # mSDRG_31
  3  2    -5.49615942E-01   # mSDRG_32
  3  3     1.77161275E+06   # mSDRG_33
#
BLOCK MSL2 Q=  1.20195104E+03  # M^2_L soft SUSY breaking masses
  1  1     2.44436035E+05   # mSLRG_11
  1  2     7.11764708E-01   # mSLRG_12
  1  3    -1.51306954E+01   # mSLRG_13
  2  1     7.11764708E-01   # mSLRG_21
  2  2     2.44428213E+05   # mSLRG_22
  2  3     7.28699498E+01   # mSLRG_23
  3  1    -1.51306954E+01   # mSLRG_31
  3  2     7.28699498E+01   # mSLRG_32
  3  3     2.41319457E+05   # mSLRG_33
#
BLOCK MSE2 Q=  1.20195104E+03  # M^2_E soft SUSY breaking masses
  1  1     9.97682891E+04   # mSERG_11
  1  2     6.08801261E-09   # mSERG_12
  1  3     1.01423906E-05   # mSERG_13
  2  1     6.08801261E-09   # mSERG_21
  2  2     9.97589360E+04   # mSERG_22
  2  3     1.15378509E-03   # mSERG_23
  3  1     1.01423906E-05   # mSERG_31
  3  2     1.15378509E-03   # mSERG_32
  3  3     9.70150720E+04   # mSERG_33
#
BLOCK SuSeFLAVLOWENERGY   #  PARAMETERS 
         1     3.39444376E-03   # Delta rho parameter
         2     5.11182018E-10   # g_mu - 2
         3     3.73074715E-04   # Br(b -> s gamma)
         4     1.70603079E-13   # Br(tau -> mu gamma)
         5     7.32782350E-15   # Br(tau -> e gamma)
         6     9.20566951E-17   # Br(mu -> e gamma)
         7     5.43318338E-16   # Br(tau -> mu mu mu)
         8     8.82689326E-17   # Br(tau -> e e e)
         9     6.80189300E-19   # Br(mu -> e e e)
#
BLOCK FINETUNE  #
         1     1.75230800E+02   # delta mZ^2/mZ^2 (mu^2)
         2     8.76436259E+01   # delta mt/mt (mu^2)
\end{verbatim}

\noindent The program prints {\tt BLOCK SELMIX} in the {\tt SLHA} output, which contains the elements of the tree level $6 \times 6$ slepton mixing matrix. The notation and the method of diagonalization is explained in more detail in the \ref{app:tree}.

\subsubsection{{\bf Non-{\tt SLHA} Interface}} 
The source file for non-{\tt SLHA} input is {\tt runonce.f}. To use the traditional \SUSEFLAV\ interface the user should execute {\tt suseflav} by modifying the corresponding input file.
This executable takes the following input files, {\tt sinputs.in}  for mSUGRA/CMSSM models, {\tt sinputs-gmsb.in} for GMSB models, {\tt sinputs-nuhm.in} for non-universal higgs model (NUHM2) models and {\tt sinputs-cnum.in} for complete non-universal models. For example, to run the program with complete non-universal boundary conditions type the following command in the terminal \\
{\tt ./suseflav <sinputs-cnum.in}\\
The sample input/output below corresponds to {\tt 10.4.4} of \cite{AbdusSalam:2011fc} with Type-I see-saw mechanism with minimal mixing in the context of mSUGRA/CMSSM scenario. To execute the program for the sample point type the following command in the terminal

{\tt ./suseflav <sinputs.in}

\begin{verbatim}
  mSUG           # MODEL name
  1              # prinstat (1 for printing output on terminal, 0 otherwise)
  1.00000E+01    # tanbeta
  5.00000E+02    # m0
  0.00000E+00    # a0
  1.05000E+03    # M1/2
  1.00000E+00    # sgn(mu)
  1.16639E-05    # G_fermi
  1.27934E+02    # alpha_em^(-1)
  1.17200E-01    # alpha_strong
  9.11876E+01    # mz_pole
  2              # one loop or two loops
  1              # quark mixing
  1              # 1= rhn on; 0 = rhn off
  CKM            # case: CKM/MNS/USD
  1.00000E-03    # spectrum tolerance
  1.72700E+02    # Mtpole
  4.23000E+00    # Mbpole
  1.77000E+00    # Mtaupole
  1.00000E+06    # MR1
  1.00000E+09    # MR2
  1.00000E+14    # MR3
  0.00000E+00    # Ynu(1,1)
  0.00000E+00    # Ynu(1,2)
  0.00000E+00    # Ynu(1,3)
  0.00000E+00    # Ynu(2,1)
  0.00000E+00    # Ynu(2,2)
  0.00000E+00    # Ynu(2,3)
  0.00000E+00    # Ynu(3,1)
  0.00000E+00    # Ynu(3,2)
  0.00000E+00    # Ynu(3,3)
  ------------------------------------------------
\end{verbatim}

\noindent The Output in \SUSEFLAV\ format is contained in the file {\tt suseflav.out}.

\begin{verbatim}
 ******** Begin Program SuSeFLAV ************
           model =    mSUG
            loop =    2
         tanbeta =    1.0000E+01
             m0  =    5.0000E+02
             a0  =    0.0000E+00
             M12 =    1.0500E+03
             m10 =    5.0000E+02
             m20 =    5.0000E+02
         sign mu =    1.0000E+00
            qmix =    1
   top pole mass =    1.7270E+02
             rhn =    1
            case =    CKM
             MR1 =    1.0000E+06
             MR2 =    1.0000E+09
             MR3 =    1.0000E+14
 ********************************************
 
  up-type yukawa at high energy:
     2.6401E-06    6.1056E-07    1.2205E-08
    -6.1068E-04    2.6401E-03    1.1119E-04
     3.8672E-03   -1.8624E-02    4.5212E-01
 
  down-type yukawa at high energy:
     7.9987E-05   -3.1319E-09    7.4082E-08
    -6.2638E-08    1.6000E-03   -7.1355E-06
     3.7936E-05   -1.8270E-04    4.5575E-02
 
  lepton-type yukawa at high energy:
     1.7539E-05   -1.3061E-10    3.1014E-09
    -2.7282E-08    3.6638E-03   -3.1200E-06
     1.1175E-05   -5.3817E-05    6.4558E-02
 
  neutrino yukawa at high energy:
     3.1809E-06    7.3564E-07    1.4835E-08
    -7.3579E-04    3.1810E-03    1.3396E-04
     4.7106E-03   -2.2686E-02    5.5081E-01
 
  up-type yukawa at msusy :
     6.3637E-06    1.4719E-06    2.3911E-08
    -1.4720E-03    6.3637E-03    2.6931E-04
     7.1199E-03   -3.4289E-02    8.3124E-01
 
  down-type yukawa at msusy :
     2.4935E-04   -1.2210E-09    2.8477E-08
    -2.4420E-08    4.9870E-03   -2.7429E-06
     1.4571E-05   -7.0171E-05    1.2939E-01
 
  lepton-type yukawa at msusy :
     2.7626E-05   -1.0563E-11    3.7615E-10
    -2.2066E-09    5.7706E-03   -3.7838E-07
     1.3609E-06   -6.5535E-06    9.9328E-02
 
  Spectrum at msusy, q =   1.7679E+03
              alpha_1  =   3.6366E-01
              alpha_2  =   6.4050E-01
              alpha_3  =   1.0297E+00
               vev1    =   2.5253E+01
               vev2    =   2.4230E+02
           newtbeta    =   9.5948E+00
                  \mu  =   1.2105E+03
              ~gluino  =   2.3009E+03
               ~stop_1 =   1.6702E+03
               ~stop_2 =   1.9889E+03
             ~scharm_R =   2.0550E+03
             ~scharm_L =   2.1380E+03
                ~sup_R =   2.0550E+03
                ~sup_L =   2.1383E+03
           ~sbottom_1  =   1.9606E+03
           ~sbottom_2  =   2.0368E+03
           ~sstrange_R =   2.0460E+03
           ~sstrange_L =   2.1394E+03
             ~sdown_R  =   2.0461E+03
             ~sdown_L  =   2.1396E+03
               ~stau_1 =   6.3321E+02
               ~stau_2 =   8.5289E+02
               ~smu_R  =   6.4202E+02
               ~smu_L  =   8.6003E+02
              ~sel_R   =   6.4206E+02
              ~sel_L   =   8.6012E+02
              ~tausnu  =   8.4727E+02
                ~musnu =   8.5526E+02
                ~elsnu =   8.5520E+02
 
 Higgs Spectrum
                 mA0   =   1.4726E+03
            mh_charged =   1.4754E+03
                 mh0   =   1.1866E+02
                  mH   =   1.4730E+03
            \alpha_h   =  -1.0481E-01
 
 Neutralino spectrum
                  N1   =   4.4421E+02
                  N2   =   8.3081E+02
                  N3   =   1.2193E+03
                  N4   =   1.2280E+03
 
 Chargino spectrum
                  C1   =   8.1458E+02
                  C2   =   1.2179E+03
 
  Low Energy Observables 
   Fine tuning, Cmz^2mu^2   =    3.5356E+02
   Fine tuning, Cmt^2mu^2   =    1.7680E+02
        Br(B  => s,gamma)   =    3.8430E-04
       Br(mu  => e,gamma)   =    3.1044E-06 X 10^(-11)
       Br(tau => mu,gamma)  =    5.9654E-06 X 10^(-08)
       Br(tau => e,gamma)   =    2.5629E-08 X 10^(-07)
         Br(tau => e,e,e)   =    3.0580E-10 X 10^(-07)
      Br(tau => mu,mu,mu)   =    1.8321E-09 X 10^(-07)
          Br(mu => e,e,e)   =    2.2577E-07 X 10^(-12)
        Br(mu => e in Ti)   =    2.0425E-06 X 10^(-12)
            (g_mu - 2)      =    1.8491E+00 X 10^(-10)

 -----------------------------

 flag for the point is  AOK                                                                                                
 --------------------------------------------------
 FLAGS AND THEIR MEANINGS
 AOK = Everything is fine.
 BMUNEG = B_mu is negative at Msusy.
 REWSB = |\mu|^2 < 0 at Msusy.
 MUNOC = Non-convergent |\mu| at Msusy.
 SW2NOC = Non-convergent Sin^2_thetaw at Mz.
 TACSPEC = Spectrum is tachyonic at Msusy.
 NPERTYUK = Non-perturbative yukawa.
 TACSPECMZ = Spectrum is tachyonic at Mz.
 FSNC = Final spectrum non-convergent.
 TACMh = Lightest CP-even neutral higgs tachyonic.
 TACMH = Heaviest CP-even neutral higgs tachyonic.
 TACMA = CP-odd neutral higgs tachyonic.
 TACMHCH = Charged higgs tachyonic.
 VARUNDER = Stepsize is zero while integrating the RGEs.
 TACSUP = SUP sector tachyonic.
 TACSDN = SDOWN sector tachyonic.
 TACSLP = SLEPTON sector tachyonic.
 TACSNU = SNEUTRINO sector tachyonic.
 LEPH = Lightest higgs mass below LEP limit.
 LEPC = Lightest chargino mass < 103.5 GeV.
 LSPSTAU = Lightest stau is LSP.
 --------------------------------------------------
\end{verbatim}


\section{Outlook}
The MEG experiment has recently reported the latest limits on the rare leptonic decay $\mu \to e + \gamma$ \cite{meg11}. Simultaneously, there are results
from LHC as well as direct detection dark matter experiments. \SUSEFLAV\ is designed to compute  supersymmetric spectrum with full flavor violation
and further to compute most of the leptonic flavor violating observables. The program can be coupled to Dark Matter routines such as \MICROMEGAS\ \cite{Belanger:2008sj}, \DARKSUSY\ \cite{Gondolo:2004sc} and \SUPERISO\ \cite{superiso} to compute the relic density and direct detection rates. The program is free under the GNU Public License and can be downloaded from the following websites:
\begin{itemize}
\item \url{http://cts.iisc.ernet.in/Suseflav/main.html}
\item \url{http://projects.hepforge.org/suseflav/}
\end{itemize}
In the present version of the program only Type-I seesaw mechanism is implemented. Future versions will include other seesaw mechanisms as well as other improvements like inclusion new flavor violating decays rates in the quark sector. 

\section*{Acknowledgments}

We appreciate discussions with and acknowledge suggestions  from L. Calibbi, S. Kraml, W. Porod  and  O. Vives.  We thank D. Grellscheid, T. Hahn and P. Slavich
for important inputs. We thank  U. Chattopadhyaya, Aseshkrishna Dutta and M. Guchait for encouragement. SKV thanks M. Ciuchini, A. Faccia, A. Masiero, P. Paradisi and L. Silvestrini for collaborations during which parts of this program were written. We thank Ketan Patel, Brandon Murakami for reporting bugs. SKV acknowledges support from DST project ``Complementarity between direct and indirect searches for Supersymmetry'' and also support from DST Ramanujan Fellowship SR/S2/RJN-25/2008. RG acknowledges support from SR/S2/RJN-25/2008. DC acknowledges partial support from SR/S2/RJN-25/2008. 


\appendix
\section{Tree-Level Masses}
\label{app:tree}
Here we suppress any gauge indices and follow the notation of BPMZ\footnote{Except in our case the sign of $\mu$ parameter is opposite of BPMZ.} \cite{Pierce:1996zz} closely. The Lagrangian contains the neutralino mass matrix as \beq{ -\frac{1}{2}{\tilde{\psi}^{0\,T}} {\cal M}_{\tilde\psi^0} \tilde{\psi}^0 + h.c. \, , }\eeq where
$\tilde\psi^0 =$ $(\widetilde B,$ $\widetilde W^0,$ $\widetilde H_d,$ $\widetilde H_u)^T$ and
\begin{equation}
{\cal M}_{\tilde\psi^0} \ =\ \left(\begin{array}{cccc} M_1 & 0 &
-M_Zc_\beta s_W & M_Zs_\beta s_W \\ 0 & M_2 & M_Zc_\beta c_W &
-M_Zs_\beta c_W \\ -M_Zc_\beta s_W & M_Zc_\beta c_W & 0 & -\mu \\
M_Zs_\beta s_W & -M_Zs_\beta c_W & -\mu & 0
\end{array} \right). \label{mchi0}
\end{equation}
We use the letter $s$ and $c$ for sine and cosine, so that
$s_\beta\equiv\sin\beta,\ c_{\beta}\equiv\cos\beta$ and $s_W (c_W)$ is
the sine (cosine) of the weak mixing angle. The $4 \times 4$ neutralino mixing matrix is an orthogonal matrix $O$ with real entries, such that $O^T {\cal M}_{\widetilde\psi^0} O$ is diagonal. The neutralinos $\chi^0_i$ are defined such that their absolute masses increase with increasing $i$. Some of their mass values can be negative.

We make the identification
${\widetilde W}^\pm = ({\widetilde W^1} \mp i{\widetilde W^2} ) / \sqrt{2}$ for the charged
winos and ${\widetilde H_u^+}, {\widetilde H_d^-}$ for the charged higgsinos.
The Lagrangian contains the chargino mass matrix as
\begin{equation}
-\half {\widetilde\psi^{-\,T}}{\cal M}_{\widetilde\psi^+}\widetilde\psi^+ + h.c. ,
\end{equation}
where~$\widetilde\psi^+ = (\widetilde W^+,\ \widetilde H_u^+)^T,\ \widetilde\psi^-=
(\widetilde W^-,\ \widetilde H_d^-)^T$ and
\begin{equation}
{\cal M}_{\tilde\psi^+}\ =\ \left( \begin{array}{cc} M_2 &
\sqrt2\,M_Ws_\beta\\\sqrt2\,M_Wc_\beta & \mu\end{array}\right).
\label{mchi}
\end{equation}
The chargino masses are found by acting on the matrix ${\cal
M}_{\tilde\psi^+}$ with a bi-unitary transformation, so that $U^*{\cal
M}_{\tilde\psi^+}V^\dagger$ is a diagonal matrix containing the two
chargino mass eigenvalues, $m_{\tilde\chi_i^+}$. The matrices $U$ and
$V$ are easily found, as they diagonalize, respectively, the matrices
${\cal M}_{\tilde\psi^+}^*{\cal M}_{\tilde\psi^+}^{\rm T}$ and ${\cal
M}_{\tilde\psi^+}^\dagger{\cal M}_{\tilde\psi^+}$. And we have taken the $U$ and $V$ matrices to be real.


The tree-level squark and slepton mass squared values for the family $i$
are found by diagonalizing the following mass
matrices ${\mathcal M}_{\tilde f}^2$ defined in the $({\tilde f}_{iL}, {\tilde
f}_{iR})^T$ basis:
\begin{equation}
\label{sqmu}
\left(\begin{array}{cc}
(m_{\tilde Q}^2)_{ii} + m_{u_i}^2 + \left(\half - \frac{2}{3} s_W^2\right)M_Z^2c_{2\beta} &
m_{u_i}\left(({\tilde A}_{\tilde u})_{ii}-\mu\cot\beta\right)\\
m_{u_i}\left(({\tilde A}_{\tilde u})_{ii}-\mu\cot\beta\right) & (m^2_{\tilde u})_{ii} +
m_{u_i}^2 +
\frac{2}{3} s_W^2 M_Z^2c_{2\beta}
\end{array}\right)\ ,
\end{equation}
\begin{equation}
\label{sqmd}
\left(\begin{array}{cc}
(m_{\tilde Q}^2)_{ii} + m_{d_i}^2 - \left(\half - \frac{1}{3} s_W^2\right)M_Z^2c_{2\beta} &
m_{d_i}\left(({\tilde A}_{\tilde d})_{ii}-\mu\tan\beta\right)\\
m_{d_i}\left(({\tilde A}_{\tilde d})_{ii}-\mu\tan\beta\right) & (m^2_{\tilde d})_{ii} +
m_{d_i}^2 -\frac{1}{3} s_W^2 M_Z^2c_{2\beta}
\end{array}\right)\ ,
\end{equation}
\begin{equation}
\label{sqme}
\Biggl(\begin{array}{cc}
(m_{\tilde L}^2)_{ii} + m_{e_i}^2 - \left(\half - s_W^2\right) M_Z^2 c_{2\beta} &
m_{e_i}\left(({\tilde A}_{\tilde e})_{ii}-\mu\tan\beta\right)\\
m_{e_i}\left(({\tilde A}_{\tilde e})_{ii}-\mu\tan\beta\right) & (m^2_{\tilde e})_{ii} +
m_{e_i}^2 - s_W^2 M_Z^2c_{2\beta}
\end{array}\Biggr)\ ,
\end{equation}
where, $m_f, e_f$ are the mass and electric charge of fermion $f$ respectively.
The mixing of the first two families is suppressed by a small fermion mass,
which we approximate to zero. The sfermion mass eigenstates are given by
\begin{equation}
\left(\begin{array}{cc} m^2_{\tilde f_1} & 0 \\ 0 & m^2_{\tilde f_2}
\end{array} \right) =
\left(\begin{array}{cc} c_f & s_f\\-s_f & c_f \end{array}\right)
{\mathcal M}_{\tilde f} ^2
\left(\begin{array}{cc} c_f & -s_f\\s_f & c_f \end{array}\right)
\end{equation}
where $c_f$ is the cosine of the sfermion mixing angle, $\cos\theta_f$, and $s_f$ is $\sin\theta_f$. These angles are given by
\begin{eqnarray}
\tan(2\theta_u) &=& \frac{2\,m_u\left({\tilde A}_{\tilde u} - \mu\cot\beta\right)}
{m_{\tilde Q}^2-m_{\tilde u}^2 + \left({\frac{1}{2}}- 2\,e_{u}\, s_W^2\right)M_Z^2 c_{2\beta}}\ ,\\
\tan(2\theta_d) &=& \frac{2\,m_d\left({\tilde A}_{\tilde d} - \mu\tan\beta\right)}
{m_{\tilde Q}^2-m_{\tilde d}^2 + \left(-{\frac{1}{2}}-2\,e_{d}\, s_W^2\right)M_Z^2c_{2\beta}}\ .
\end{eqnarray}

To calculate the lepton flavor violating decays, we diagonalize the full $6\times6$ sleptonic mass matrix (${\cal M}^{2}_{\tilde l}$) by $U^{T}\ {\cal M}^{2}_{\tilde l}\ U$, where $U$ is the sleptonic mixing matrix with real entries. In the gauge basis of $\left\{\tilde{e}_L,\tilde{\mu}_L,\tilde{\tau}_L,\tilde{e}_R,\tilde{\mu}_R,\tilde{\tau}_R\right\}^{T}$ the sleptonic mass matrix is defined as 
\begin{equation}
{\cal M}^{2}_{\tilde l}\ =\
\begin{pmatrix}
\m_{\tilde L}^2 + \left[m_{e_i}^2 - \left(\half - s_W^2\right) M_Z^2 c_{2\beta}\right]{\bf 1} &
m_{e_i}\left({\tilde \A}_{\tilde e} - \mu \tan\beta \; {\bf 1} \right)\\
m_{e_i}\left({\tilde \A}^{T}_{\tilde e} - \mu \tan\beta \ {\bf 1} \right) & \m^2_{\tilde e} +
\left[m_{e_i}^2 - s_W^2 M_Z^2c_{2\beta}\right] {\bf 1}
\end{pmatrix}\ ,
\end{equation}
where $\m_{\tilde L}^2$, $\m^2_{\tilde e}$ and ${\tilde \A}_{\tilde e}$ are the scalar mass matrices and tri-linear coupling matrix respectively. Using the full $6\times6$ sleptonic mixing matrix we define the lepton flavor violating couplings following Hisano et al. \cite{Hisano}. With these couplings we calculate the rare lepton flavor violating decays as described in section \ref{observables}.


Given values for $\tan\beta$ one can write the CP-odd Higgs-boson mass, $m_A$, and the other Higgs masses at tree level by
\begin{equation}
\label{trmA}
m_A^2 = \frac{2B_\mu}{\sin2\beta} = 2|\mu|^2 + m^2_{H_u} + m^2_{H_d}\ ,
\end{equation}
\begin{equation}
\label{trmh}
m_{H,h}^2 \ =\ \frac{1}{2}\Biggl(m_A^2 + M_Z^2 \pm \sqrt{\left(m_A^2 +
M_Z^2 \right)^2 - 4 m_A^2 M_Z^2 c_{2\beta}^2}\Biggr)\ ,
\end{equation}
and %
\begin{equation}
\label{trmhc}
m_{H^\pm}^2 \ =\ m_A^2\ +\ M_W^2\ .
\end{equation}
The CP-even gauge eigenstates $(H^0_d,\, H^0_u)$ are rotated by the angle
$\alpha$ into the mass eigenstates $(H,\, h)$ as follows,
\begin{equation}
\frac{1}{\sqrt{2}}\left(\begin{array}{c}H\\[2mm]h\end{array}\right) \ =
\ \left(\begin{array}{cc} c_\alpha & s_\alpha\\[2mm]-s_\alpha & c_\alpha
\end{array}\right)\left(\begin{array}{c}\mathfrak{Re} H^0_d\\[2mm]\mathfrak{Re} H^0_u\end{array}\right)~.
\label{rotate h}
\end{equation}
At tree level, the angle $\alpha$ is given by
\begin{equation}
\tan2\alpha \ =\ \frac{m_A^2 + M_Z^2}{m_A^2-M_Z^2}\tan2\beta\ .
\end{equation}
\section{One Loop Threshold Corrections}
\label{oneloop}
We compute flavor conserving complete one-loop corrections to masses and couplings in MSSM following BPMZ \cite{Pierce:1996zz}. In this appendix we present the summary of one-loop corrections implemented in \SUSEFLAV.

\begin{itemize}
\item {\bf Quarks and Leptons}

The corrections to fermions are evaluated at the weak scale. The running masses $m_{\hat{f
}}$ are related to the corresponding $\overline{DR}$  masses by the expression
\begin{equation}
\label{qandl}
m_f = m_{\hat{f}} - \Sigma_f^{BPMZ} (m^2_{f})
\end{equation}
Where, $\Sigma_f^{BPMZ} (m^2_{f})$ is the one loop self-energy of the fermion $f$.
We follow equation D.18 of BPMZ \cite{Pierce:1996zz}. At the present version of \SUSEFLAV\ we add the correction only to the third generation fermions.

\item{\bf W and Z  Bosons}

Corrections to W and Z bosons are evaluated at the weak scale as well as EWSB scale. The $\overline{DR}$ running W and Z boson mass squared at the scale $Q$ ($\hat M_Z^2(Q)$ and $\hat M_W^2(Q)$) are related to the physical pole mass of gauge bosons as follows
\begin{eqnarray}
M_Z^2  &=& \hat M_Z^2(Q) \ - \ \Pi^T_{ZZ}(M_Z^2) \label{mz} \\  
M_W^2  &=& \hat M_W^2(Q) \ -  \ \Pi^T_{WW}(M_W^2)
\end{eqnarray}

Where, $\Pi^T_{ZZ}(M_Z^2)$ and $\Pi^T_{WW}(M_W^2)$ are the transverse part of  self energy terms. Consult appendix D of BPMZ \cite{Pierce:1996zz} for detailed discussion and the complete expression of the self energy terms.

\item{\bf sQuarks and sLeptons}

Flavor conserving one-loop masses and mixings for squarks and sleptons are evaluated by diagonalizing ${\cal M}_{\tilde f}^2(p^2)$ for eigenvalues at the EWSB scale. 
\begin{equation}
{\cal M}_{\tilde f}^2(p^2)\ =\ \left(\begin{array}{cc} M_{\tilde
f_L\tilde f_L}^2-\Pi_{\tilde f_L\tilde f_L}(p^2) & \qquad
M_{\tilde f_L\tilde f_R}^2-\Pi_{\tilde f_L\tilde f_R}(p^2)
\\[2mm] M_{\tilde f_R\tilde f_L}^2-\Pi_{\tilde f_R\tilde
f_L}(p^2) & \qquad M_{\tilde f_R\tilde f_R}^2-\Pi_{\tilde
f_R\tilde f_R}(p^2)
\end{array}\right)~
\end{equation}
$M^2_{\tilde{f} \tilde{f}}$ is the $2\times2$ tree level mass matrix defined in equations \ref{sqmu}, \ref{sqmd} and \ref{sqme}. The self-energy terms of the above matrix \Big($\Pi_{\tilde{f}_L\tilde{f}_L}$, $\Pi_{\tilde{f}_L\tilde{f}_R}$, $\Pi_{\tilde{f}_R\tilde{f}_L}$, $\Pi_{\tilde{f}_R\tilde{f}_R}$\Big) are defined in appendix D of BPMZ \cite{Pierce:1996zz}.

\item {\bf Higgs Bosons}

The mass of the two loop CP-even higges are obtained by diagonalizing ${\cal M}^2_s(p^2)$ at the EWSB scale. Where the matrix ${\cal M}^2_s(p^2)$ is given by,
\begin{equation}
{\cal M}^2_s(p^2) \ =\ \left(\begin{array}{cc} \hat M_Z^2c_\beta^2 +
\hat m_A^2s_\beta^2 -\,\Pi_{s_1s_1}(p^2) + t_1/v_1 & -(\hat
M_Z^2 + \hat m_A^2)s_\beta c_\beta -\,\Pi_{s_1s_2}(p^2)\\[2mm]
-(\hat M_Z^2 + \hat m_A^2)s_\beta c_\beta -\,\Pi_{s_2s_1}(p^2)
& \hat M_Z^2s_\beta^2 + \hat m_A^2c_\beta^2 -\,\Pi_{s_2s_2}(p^2)
+ t_2/v_2
\end{array}\right)\ 
\end{equation}
The self energy terms $\Pi_{s_1 s_1},\,\Pi_{s_1 s_2},\,\Pi_{s_2 s_1}$ and $\Pi_{s_2 s_2}$ are presented in appendix D of BPMZ \cite{Pierce:1996zz}. Whereas, one loop tadpole contributions $t_1/v_1$ and $t_2/v_2$ are presented in appendix E of BPMZ \cite{Pierce:1996zz}. Besides the complete one loop correction to $m_h$ and $m_{H}$, we evaluate the top mass enhanced dominant two loop corrections provided in \cite{Heinemeyer:1999be}. From version 1.2 onwards, we have implemented the full two loop Higgs corrections due to Slavich \textit{et. al} \cite{pietro}. 

The tree level mass of CP-odd higgs boson, $m_A$  and charged higgs boson $m_{H^{\pm}}$ is given by equations \ref{trmA} and \ref{trmhc} respectively.
The one loop correction to  masses is evaluated at EWSB scale and  given by
\begin{eqnarray}
 \hat{m}^2_A &=&  m^2_A + \Pi_{AA}(m_A^2) - b_A \label{1lma} \\ \hat{m}_{H^{\pm}} &=& m^2_A + M^2_{W} +\Pi_{AA}(m_A^2)  +  \Pi^T_{WW}(M_W^2) - \Pi_{H^+ H^-}(m^2_{H^{\pm}})\label{1lmhc} 
\end{eqnarray}
Where, $b_A = s^2_{\beta}\ t_1/v_1 + c^2_{\beta}\ t_2/v_2$.

\item{\bf Neutralino and Chargino}

The one loop corrected neutralino mass matrix has the following form,
\begin{equation}
 {\cal M}_{\tilde\psi^0}\ +\ \frac{1}{2}\left(\delta {\cal
M}_{\tilde\psi^0}(p^2) + \delta{\cal M}_{\tilde\psi^0}^T(p^2)\right)~
\label{chi01}
\end{equation}
where
\begin{equation}
 \delta{\cal M}_{\tilde\psi^0}(p^2) \ =
\ -\ \Sigma_R^0(p^2){\cal M}_{\tilde\psi^0}
\ -\ {\cal M}_{\tilde\psi^0}\Sigma_L^0(p^2)
\ -\ \Sigma_S^0(p^2)\ 
\label{chi02}
\end{equation}
Where, ${\cal M}_{\tilde\psi^0}$ is the tree level neutralino mass matrix defined in \ref{mchi0}. And the matrix $\delta{\cal M}_{\tilde\psi^0}(p^2)$ is the one loop correction to the neutralino mass matrix.

The one-loop chargino mass matrix is as follows,
\begin{equation}
{\cal M}_{\tilde\psi^+}\ -\ \Sigma_R^+(p^2)\,{\cal M}_{\tilde\psi^+}
\ -\ {\cal M}_{\tilde\psi^+}\,\Sigma_L^+(p^2)\ -\ \Sigma_S^+(p^2)~
\label{chic1}
\end{equation}
Where, ${\cal M}_{\tilde\psi^+}$ is the tree level chargino mass matrix defined in \ref{mchi}. We evaluate the self energies $\Sigma^{+,0}_{L,R,S}$ with $p^2 = Q$(EWSB scale). See appendix D of BPMZ for the complete expressions of these self-energies. One loop neutralino and chargino mass matrices are then diagonalized to obtain the eigenvalues and 
eigenvectors at EWSB scale.

\item{\bf Gauge Couplings and $\sin^2\theta_W$}

The correction to $\overline{DR}$ electromagnetic coupling at the weak scale is given by,
\begin{equation}
\label{alphem}
\hat{\alpha} = \frac{\alpha_{em}}{1 - \Delta \hat{\alpha}}  \qquad {\rm and} \qquad\alpha_{em} = \frac{1}{127.934}   
\end{equation}
Where $\Delta \hat{\alpha}$ contains contribution from SM particles as well as SUSY particles. Similarly, the correction to strong coupling $\alpha_s$ at the weak scale is given by,
\begin{equation}
\label{alphas}
\hat{\alpha}_s = \frac{\alpha_s}{1 - \Delta\alpha_s}
\end{equation}
$\Delta\alpha_s$ receives contribution from top quark, gluino and squarks. The exact expression given by equation 3 of BPMZ \cite{Pierce:1996zz}. The correction to $\overline{DR}$ weak mixing angle is given by,
\begin{equation}
\label{weakmix}
\sin^2 2\hat{\theta}_W = \frac{4 \pi \hat{\alpha}}{\sqrt{2} M_Z^2 G_{\mu} (1 - \Delta \hat{r})}
\end{equation}
The exact expression of $\Delta \hat{\alpha}$ and  $\Delta \hat{r}$ is presented in the appendix C of BPMZ \cite{Pierce:1996zz}.  

\item{\bf Gluino}

One loop corrected gluino  mass $m_{\tilde{g}}$ at scale Q  is given by,
\begin{equation}
\label{gluino}
m_{\tilde{g}} = M_3(Q) - \Sigma_{\tilde{g}} (m_{\tilde{g}})
\end{equation}
where $M_3(Q)$ is the tree level gluino mass generated by the RGEs and $\Sigma_{\tilde{g}} (m_{\tilde{g}})$ is the one loop self-energy contribution to the gluino mass. See equation D.44 of BPMZ \cite{Pierce:1996zz} for the complete expression.

\end{itemize}

\section{Program File Structure}
\label{prog_flow}
\begin{figure}[htb]
\begin{center}
 \includegraphics[width=0.90\textwidth,angle=0]{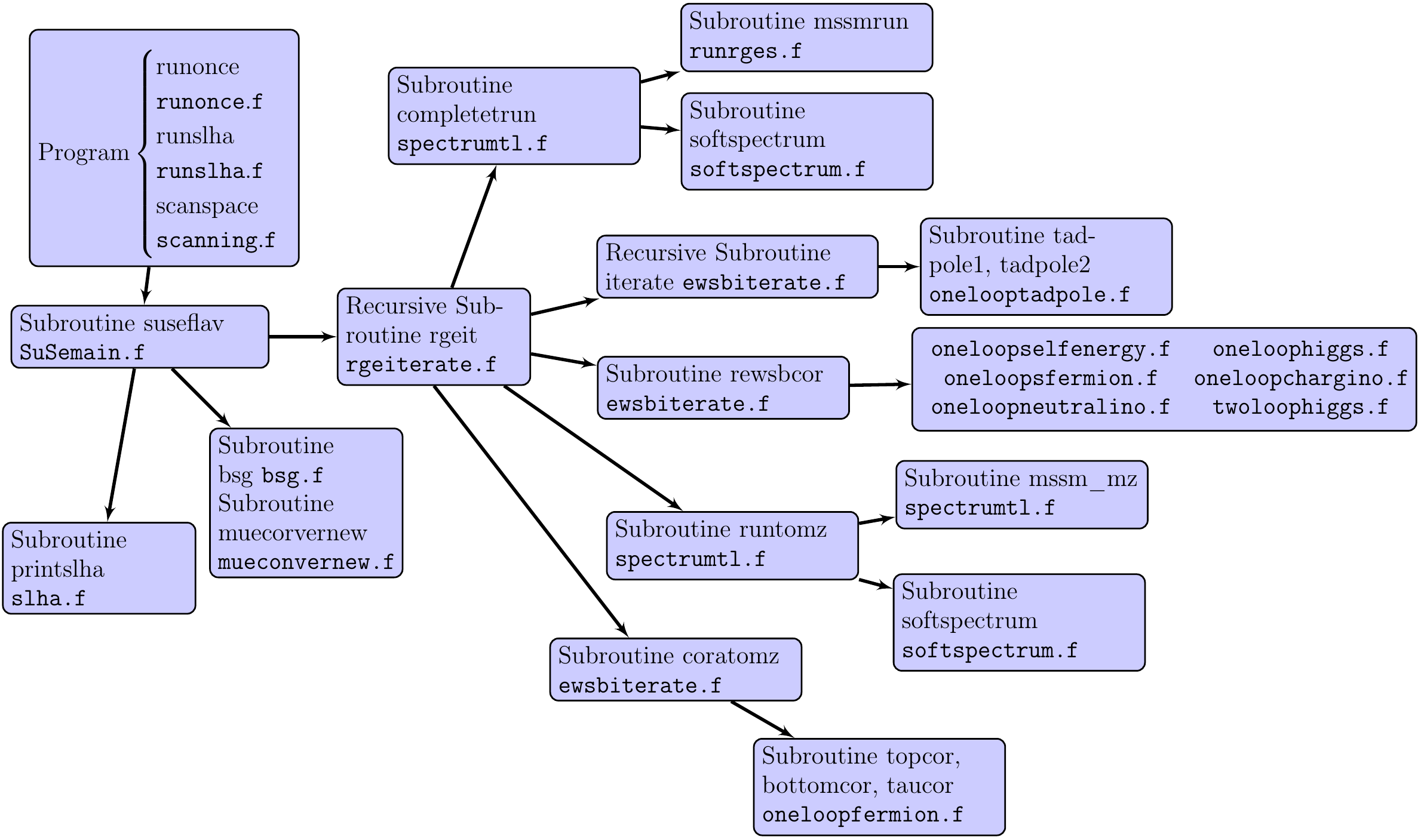}
\caption{File structure and program flow }
\label{programflow}
\end{center}
\end{figure}

In this appendix we briefly list the subroutines of interest in the program and its purpose (for more details see the technical manual provided with the program). Fig. (\ref{programflow}) depicts the program structure and the flow of variables in the program. The subroutine name is printed first followed by the respective file which contains the subroutine.

The program begins by reading inputs from the main program contained in the files `\texttt{runonce.f}', `\texttt{runslha.f}' (for {\tt SLHA} input format), `\texttt{scanning.f}'.  The convention followed in the program is such that all the input parameters which do not receive one-loop susy threshold correction such as the masses of the first two generation of quarks and leptons are contained in the file `\texttt{stdinputs.h}'. Whereas, parameters such as $\sin^2 \theta_W$, gauge couplings and the third generation quark and lepton masses which receive one-loop SUSY threshold correction are contained in the file `\texttt{SuSemain.f}'. Moreover, the complete $3 \times 3$ structure of the Yukawas and soft terms are defined in `\texttt{SuSemain.f}'.

\begin{itemize}
\item \texttt {rgeiterate.f:} The heart of the program. The iterative algorithm is contained here.

\item \texttt{softspectrum.f:} Given the RGE output at low scale, this routine computes the complete tree level SUSY spectrum by diagonalizing the mass matrices. 

\item \texttt{slha.f:} Contains the ingredients for {\tt SLHA} input/output interface.

\item \texttt {mueconvernew.f:} Contains routines which calculate the decay rates and branching fractions of rare lepton flavor violating decays.

\item \texttt {mssmrge.f:} The complete MSSM two loop RGEs are contained in this file.

\item \texttt{oneloop}\textit{particle}\texttt{.f:} A set of several files which contain the one loop corrections to the corresponding \textit{particle}, which could be gauge, neutralino, charging, sfermion etc.

\item \texttt {oneloopPV.f:} The analytical form of all scalar one loop Passarino-Veltman functions are contained in this file.

\item \texttt{math.f:} Routines for matrix manipulations,  integration routine for ordinary differential equations using  Runge-Kutta with adaptive step size and other general purpose functions such as random number generator.

\end{itemize}

\section{Spectrum Comparison with Other Codes}
\label{comp}
In this section we compare the spectrum of \SUSEFLAV\ with other codes for two different parameter space points form \cite{AbdusSalam:2011fc}. For both the points we have set RHN equals to zero. 

\begin{table}[htdp]
\caption{Spectrum comparison for point {\tt 10.1.5} of \cite{AbdusSalam:2011fc}}
\begin{center}
\renewcommand{\arraystretch}{1.1}
\begin{tabular}{c c c c c}
\hline
{\tt 10.1.5} & \multicolumn{4}{c}{$m_{0} = 175$, $M_{\half} = 700$, $\tb = 10$, $A_{0} = 0$, sign$(\mu) > 0$} \\ \hline
Parameters & \SUSEFLAV & \SOFTSUSY & \SUSPECT & \spheno  \\ \hline
 $\W$ &    8.05096462E+01 & 8.03860774e+01 & 8.04813477E+01 & 8.03115948E+01 \\
 $h$ &     1.16294107E+02 & 1.16694326e+02 & 1.16799828E+02 & 1.16829168E+02 \\
 $H$ &     9.66133037E+02 & 9.74430227e+02 & 9.68302501E+02 & 9.79779724E+02 \\
 $A$ &     9.65805171E+02 & 9.74206243e+02 & 9.68101504E+02 & 9.79577186E+02 \\
 $\Hp$ &   9.69592528E+02 & 9.77728714e+02 & 9.71660981E+02 & 9.83266787E+02 \\
 $\qL$ &   1.43474627E+03 & 1.43580606e+03 & 1.43214020E+03 & 1.43992830E+03 \\
 $\qR$ &   1.37749634E+03 & 1.37825389e+03 & 1.37724979E+03 & 1.38138749E+03 \\
 $\stL$ &  1.11659996E+03 & 1.11105612e+03 & 1.11462082E+03 & 1.10949455E+03 \\
 $\stR$ &  1.35668185E+03 & 1.35225046e+03 & 1.35767238E+03 & 1.35426026E+03 \\
 $\sbL$ &  1.31765879E+03 & 1.31707770e+03 & 1.32640786E+03 & 1.31861062E+03 \\
 $\sbR$ &  1.36683066E+03 & 1.36729669e+03 & 1.36701474E+03 & 1.37028335E+03 \\
 $\seL$ &  5.01303257E+02 & 5.01757612e+02 & 4.94643201E+02 & 5.01101877E+02 \\
 $\seR$ &  3.23056643E+02 & 3.18458635e+02 & 3.13310890E+02 & 3.18342964E+02 \\
 $\stauL$ & 3.15781781E+02 & 3.11269901e+02 & 3.06660771E+02 & 3.11160940E+02 \\
 $\stauR$ & 5.01464374E+02 & 5.01745556e+02 & 4.94979107E+02 & 5.01126010E+02 \\
 $\nuL$ &  4.94057805E+02 & 4.95324112e+02 & 4.88544346E+02 & 4.94578335E+02 \\
 $\nutau$ & 4.92669260E+02 & 4.93751173e+02 & 4.87265068E+02 & 4.92997226E+02 \\
 $\tilde{g}$ & 1.56391437E+03 & 1.56476030e+03 & 1.56495608E+03 & 1.57232851E+03 \\
 $\neut{1}$ & 2.95219584E+02 & 2.91289971e+02 & 2.92926561E+02 & 2.92509855E+02 \\ 
 $\neut{2}$ & 5.45858411E+02 & 5.51079728e+02 & 5.52119478E+02 & 5.52386977E+02 \\
 $\neut{3}$ & 8.44975403E+02 & 8.50445968e+02 & 8.60389258E+02 & 8.56133860E+02 \\
 $\neut{4}$ & 8.56487466E+02 & 8.62039492e+02 & 8.72046145E+02 & 8.67579765E+02 \\
 $\charg{1}$ & 5.35344839E+02 & 5.51280182e+02 & 5.52043599E+02 & 5.52530182E+02 \\
 $\charg{2}$ & 8.49067665E+02 & 8.61668017e+02 & 8.71711677E+02 & 8.67856046E+02 \\ 
 $\mu$ & 8.37298648E+02 & 8.44864881e+02 & 8.55544693E+02 & 8.50798325E+02 \\
 $\msusy$ & 1.19242602E+03 & 1.19031414e+03 & 1.19174751E+03 & 1.22578828E+03 \\
 $\mgut$ & 1.26703749E+16 & 1.67323027e+16 & $-$ & 1.53713886E+16 \\
\hline
\end{tabular}
\end{center}
\label{comp1}
\end{table}%

\begin{table}[htdp]
\caption{Spectrum comparison for point {\tt 40.2.3} of \cite{AbdusSalam:2011fc}}
\begin{center}
\renewcommand{\arraystretch}{1.1}
\begin{tabular}{c c c c c}
\hline
 {\tt 40.2.3} & \multicolumn{4}{c}{$m_{0} = 650$, $M_{\half} = 550$, $\tb = 40$, $A_{0} = -500$, sign$(\mu) > 0$} \\ \hline
Parameters & \SUSEFLAV & \SOFTSUSY & \SUSPECT & \spheno  \\ \hline
 $\W$ &    8.05066107E+01 & 8.03827802e+01 & 8.04773725E+01 & 8.03087734E+01 \\
 $h$ &     1.16958052E+02 & 1.17297317e+02 & 1.17322254E+02 & 1.17682268E+02 \\
 $H$ &     7.30962046E+02 & 7.47689856e+02 & 7.18511292E+02 & 7.72497154E+02 \\
 $A$ &     7.31051866E+02 & 7.47643928e+02 & 7.18551942E+02 & 7.72499188E+02 \\
 $\Hp$ &   7.35889065E+02 & 7.52307728e+02 & 7.23362920E+02 & 7.77278228E+02 \\
 $\quL$ &   1.30348659E+03 & 1.28439273e+03 & 1.30388612E+03 & 1.31020706E+03 \\
 $\quR$ &   1.26813307E+03 & 1.24781930e+03 & 1.27013696E+03 & 1.27389691E+03 \\
 $\qdL$ &   1.30579237E+03 & 1.28668811e+03 & 1.30621620E+03 & 1.31246645E+03 \\
 $\qdR$ &   1.26498184E+03 & 1.24450694e+03 & 1.26672885E+03 & 1.27054542E+03 \\
 $\stL$ &  9.24983971E+02 & 9.21498284e+02 & 9.69783691E+02 & 9.19586596E+02 \\
 $\stR$ &  1.14470194E+03 & 1.14136271e+03 & 1.17704096E+03 & 1.14412090E+03 \\
 $\sbL$ &  1.08597531E+03 & 1.08516841e+03 & 1.11715522E+03 & 1.09021482E+03 \\
 $\sbR$ &  1.15972419E+03 & 1.16046164e+03 & 1.19253887E+03 & 1.16721088E+03 \\
 $\seL$ &  7.46153736E+02 & 7.46218781e+02 & 7.43808878E+02 & 7.46077658E+02 \\
 $\seR$ &  6.83925820E+02 & 6.82393714e+02 & 6.80821036E+02 & 6.82370617E+02 \\
 $\stauL$ & 5.23418984E+02 & 5.18531351e+02 & 5.54473085E+02 & 5.18371511E+02 \\
 $\stauR$ & 6.95666250E+02 & 6.94111081e+02 & 7.16257580E+02 & 6.94363534E+02 \\
 $\nuL$ &  7.40378630E+02 & 7.41675551e+02 & 7.39679081E+02 & 7.41481196E+02 \\
 $\nutau$ & 6.79971865E+02 & 6.77798789e+02 & 6.94510135E+02 & 6.77596601E+02 \\
 $\tilde{g}$ & 1.27460200E+03 & 1.27746572e+03 & 1.28080280E+03 & 1.28539752E+03 \\
 $\neut{1}$ & 2.33065416E+02 & 2.29995196e+02 & 2.31351813E+02 & 2.30719919E+02 \\ 
 $\neut{2}$ & 4.32028446E+02 & 4.38269965e+02 & 4.43044390E+02 & 4.39375323E+02 \\
 $\neut{3}$ & 7.36025983E+02 & 7.45547839e+02 & 9.13965835E+02 & 7.54411085E+02 \\
 $\neut{4}$ & 7.44572370E+02 & 7.54116707e+02 & 9.19380416E+02 & 7.62689372E+02 \\
 $\charg{1}$ & 4.22498882E+02 & 4.38347771e+02 & 4.43031333E+02 & 4.39557714E+02 \\
 $\charg{2}$ & 7.40879408E+02 & 7.54555638e+02 & 9.20007967E+02 & 7.63794652E+02 \\
 $\mu$ & 7.29597776E+02 & 7.41004045e+02 & 9.12311446E+02 & 7.50392113E+02 \\
 $\msusy$ & 9.97283261E+02 & 9.96175665e+02 & 1.03350532E+03 & 1.02572777E+03 \\
 $\mgut$ & 1.23171917E+16 & 1.88768830e+16 & $-$ & 1.75795520E+16 \\
\hline
\end{tabular}
\end{center}
\label{comp2}
\end{table}%









%
%
%

\end{document}
